\documentclass[twocolumn,showpacs,aps,prd,preprintnumbers,nofootinbib]{revtex4-1}
\usepackage{amsmath,graphicx,amssymb,slashed}
\begin{document}
\title{Polarization of top produced in particle decays in an arbitrary frame}
\author{Arunprasath V.} 
\email{arunprasath@cts.iisc.ernet.in}
\author{Rohini M. Godbole}
\email{rohini@cts.iisc.ernet.in}
\affiliation{Centre for High Energy Physics, Indian Institute of Science, 
Bengaluru, 560012}
\author{Ritesh K. Singh}
\email{ritesh.singh@iiserkol.ac.in}
\affiliation{Department of Physical Sciences, Indian Institute of Science
Education and Research Kolkata, Mohanpur, 741246, India}
\begin{abstract}
In most of the models beyond the Standard Model, the top quark is expected to 
be polarized when produced in the decay of some heavier particle, like the gluino or the stop. The polarization is constructed, in an experiment or in 
simulations, through the distribution of top decay products. Here, we propose 
an estimator of top quark polarization that depends only on the kinematics 
of it's mother particle, apart from its decay couplings to top quarks, and is 
given in terms of the top polarization expected in the rest frame of the 
decaying particle. This estimator allows one to estimate the top polarization 
without performing a full simulation. We find this estimator is independent of 
the production angle of the mother, top decay angle (for unpolarized 
mother), and the spin of the mother particle. We study the quality of the
estimator with finite width of the mother particle via examples of gluinos and
stops decaying into top quark at LHC. We also point out how for the mass 
spectra of gluinos and top squarks currently expected in a `natural' scenario, 
the polarization of the top quarks produced in the gluino decays can uniquely 
track the mixing angle in the stop sector.
\end{abstract}
\maketitle
\section{Introduction}\label{sec:1}
The top quark is the heaviest known fundamental particle of the SM. The LHC 
produces top quarks copiously, enabling a precision study of it's 
properties~\cite{Beneke:2000hk,Han:2008xb,Bernreuther:2008ju,Schilling:2012dx,
BARBERIS:2013wba,JABEEN:2013mva}. The importance of the studies of the top 
quark's properties lies not only in the validation of the SM in the top sector 
but also in probing effects of any possible new physics (NP). Since the mass of 
the top quark is close to the EW symmetry breaking scale, it is expected to play 
an important role in the electroweak symmetry breaking~\cite{Hill:2002ap,
Schmaltz:2005ky}. One of the important properties of the top quark is that it 
decays before hadronization sets in. This property makes it possible to obtain 
information on top spin through the kinematic distributions of the decay 
products~\cite{Jezabek:1988ja,Czarnecki:1990pe,Kane:1991bg}.

The polarization of the produced top quark is determined by the production 
mechanism and hence varies from process to process. For example, the 
polarization of the top in the $t\bar{t}$ production process is negligible due 
to the parity conserving nature of the strong interaction -- a purely vector 
interaction. The top polarization in $t\bar{t}$ production is about $0.4\%$ for 
14 TeV LHC, at NLO in QCD with 1-loop weak and QED corrections. This is the 
value in the so called helicity basis where the spin quantization axis is along 
the direction of motion of the top~\cite{Bernreuther:2015yna}. On the other 
hand, the weak interaction mediated top production process, the single top 
production, produces highly polarized top quarks due to the $V-A$ nature of the 
interaction. For example, in the spectator basis, where the top spin 
quantization direction is taken along the direction of the light quark jet that 
scatters away from the top quark, the single top production process produces a 
top polarization of about $-0.99$~\cite{Schwienhorst:2010je,Mahlon:1999gz}, at 
leading order. In the helicity basis, in the center of mass frame of the top 
quark and the spectator jet, the single top polarization is about $0.99$, at 
leading order~\cite{Schwienhorst:2010je,Mahlon:1999gz}. Polarized top quarks can 
also be produced in processes of various Beyond the Standard Model (BSM) 
scenarios, such as, minimal supersymmetric standard model 
(MSSM)~\cite{Perelstein:2008zt,Low:2013aza,Belanger:2013gha,Belanger:2012tm,
Shelton:2008nq}, R-parity violating MSSM~\cite{Hikasa:1999wy,Arai:2010ci}, 
warped extra dimensions~\cite{Agashe:2006hk,Choudhury:2007ux}, little 
Higgs~\cite{Godbole:2010kr} etc. Any new physics affecting the top production 
and which is chiral in nature can affect the polarization of the top. Hence, top 
polarization can be used as a signature of new physics in the top 
production~\cite{Gopalakrishna:2010xm,Godbole:2010kr,Huitu:2010ad,
Rindani:2011pk,Biswal:2012dr,Belanger:2012tm,Belanger:2013gha,
Aguilar-Saavedra:2014eqa,Cao:2010nw,Choudhury:2010cd,Krohn:2011tw,
Falkowski:2011zr,Godbole:2011vw,Baumgart:2013yra,Cao:2011hr,Fajfer:2012si,
Choudhury:2007ux}.

In many of the BSM scenarios mentioned above, the top quark can be produced 
through the decays of some heavy particle postulated 
therein~\cite{Perelstein:2008zt,Shelton:2008nq,Belanger:2013gha,
Belanger:2012tm}. In these cases, the top polarization is determined in the 
rest frame of the mother particle by dynamical parameters of the interactions 
that are responsible for the decay of the mother particle and is given by a simple analytical expression. However, in the frame 
where the top polarization is measured, laboratory frame (say), the decaying 
mother particle is not at rest, in general\footnote{Measurement of top 
polarization in the laboratory frame has the advantage that it does not require 
the reconstruction of the rest frame of the top quark. An estimation of top 
polarization that would be observed in the laboratory frame is useful in the 
construction of appropriate top spin observables. Our work illustrates some of 
the important issues in the estimation of lab frame top polarization.}. Since 
the top helicity states are not invariant under arbitrary Lorentz 
transformations the top polarization measured in the laboratory (lab) frame is 
not the same as the one given in the rest frame of the mother particle. The 
two values are related by a kinematical factor which, in general, would depend 
on the direction and magnitude of the Lorentz boost required to reach the lab 
frame (or any frame where the mother particle is moving) from the rest frame of 
the mother particle. In this work we determine the kinematic factor and provide 
its explicit analytical expression, assuming that the mother particle is unpolarized and  has a narrow width. We choose two processes where the top is produced through the decay of another particle: $pp\rightarrow \tilde{g}\tilde{g}\rightarrow \tilde{g}t\tilde{t}_1^{\ast}$, and 
$pp\rightarrow \tilde{t}_1^{\ast}\tilde{t}_1\rightarrow \tilde{t}_1^{\ast}t
\tilde{\chi}_1^{0}$. These processes can be written in a general form: 
\begin{equation}\label{eq:1}
pp\rightarrow A+A\rightarrow A+(t+B)
\end{equation}
where $A$ is the heavy particle which decays\footnote{We mostly assume that the 
unstable particle $A$ has a narrow width i.e., $\Gamma_A/M_A\ll 1$ and also 
that the masses of $A,B,t$ are widely separated, but later generalize to the 
case where the narrow width approximation is lifted for the particle $A$.} as $A\rightarrow t+B$. For these processes, we provide an estimator of top polarization $\mathcal{P}_{estimator}$ which can give an estimate of top polarization without simulating the heavy particle decay and using simply the kinematical distribution of the mother particle A. This estimator can be written as a convolution of the top polarization in the rest frame of the mother particle which is determined by the kinematic and dynamical factors mentioned above, and the velocity distribution of the mother particle in that frame. Since the velocity distribution of the mother particle is determined by the (parton distribution function) PDF factors of the $pp$ collision, this estimate of the top polarization can be understood as a weighted average of top polarization over the entire sample of events.   
\begin{equation}\label{eq:2}
\mathcal{P}_{estimator} \equiv \frac{1}{\sigma_{AA}}\int \frac{d\sigma_{AA}}
{d\beta_{A}} \ \ \mathcal{P}(\beta_{A}).
\end{equation} 
Here, $\sigma_{AA}$ is the cross section for the pair production of $A$ in the 
process: $pp\rightarrow A\bar{A}$ and $\beta_A$ is the velocity 
of $A$ in the given frame. In the following discussions we choose this frame to 
be either the lab frame or the parton center of mass (PCM) frame. We find that 
$\mathcal{P}(\beta_A)$ in any chosen frame depends not on the direction of 
emission of the unpolarized mother particle $A$ in that frame but only the 
magnitude of its velocity. We find that the estimator gives a good 
approximation of the true value of top polarization when the events are 
dominated by events where the mother particle is on-shell. When the sample of 
events is dominated by off-shell decays of the mother, a good estimation of 
top polarization can be obtained, in the case of scalar mother, by assuming 
that the mass of the mother is distributed as Breit-Wigner distribution. We 
find that this does not work very well for the case of  spin-1/2 mother 
particle and we explain the reason behind it. 



 We choose these examples of gluino/stop pair production and their decay for their phenomenological importance, as explained in the following. After the LHC discovery of a light Higgs with SM-like couplings and a mass about
125 GeV~\cite{Chatrchyan:2012xdj,Aad:2012tfa}, questions on the naturalness in 
the Higgs sector of the SM have become urgent.  Given the light Higgs mass of 
125 GeV, models within the framework of MSSM, typically require, a large stop 
mixing, stop mass eigenstates with masses $\sim 1\;TeV$, and a heavy 
gluino~\cite{Hall:2011aa,Arbey:2011ab,Baer:2012uy}. However, Higgs sector can be
also natural and be consistent with the observation of a light Higgs, in models 
where one of the two stop mass eigenstates is light ($0.5-1.0$ TeV) and the 
other one is heavy ($\sim 1.0$ TeV)~\cite{Barger:2012hr,CahillRowley:2012rv,
Drees:2015aeo}. When one of the stop mass eigenstates is light and when gluinos 
are not too heavy so that their production cross section remains accessible at 
the LHC, the corresponding model parameters can be probed at the LHC through the
polarization of top quarks produced in their decays. The top quark produced in 
the decay of a gluino or a stop is expected to be polarized because of the 
chiral nature of its coupling with the gluino or the stop, since the mass 
eigenstate $\tilde{t}_1$ (say) is an arbitrary mixture of $\tilde{t}_L$ and 
$\tilde{t}_R$ and the neutralino $\tilde{\chi}_1^0$ is an arbitrary mixture of 
higgsino and gaugino.

In the decays of a gluino where a top is produced, $\tilde{g}\rightarrow 
\tilde{t}_1^{\ast}t$ and $\tilde{g}\rightarrow \tilde{t}_2^{\ast}t$, the 
polarization of the top produced is a direct measure of the stop mixing angles,
as we shall show in Section \ref{sec:2}. On the other hand, when the top is 
produced from the decay of a stop, the top polarization in the stop rest frame 
depends not only on the stop mixing angle but also on gaugino and higgsino 
content of the neutralino~\cite{Nojiri:1994it,Perelstein:2008zt,
Belanger:2012tm,Kitano:2002ss}. Hence, it is interesting to calculate top 
polarization in gluino decays though the cross sections may be smaller as the 
limits on gluino masses have already touched~$\sim$~TeV. The stop decays, on 
the other hand, can have higher cross sections as the LHC data allows them to 
be much lighter compared to a gluino, though the top polarization now depends 
additionally on parameters such as mixing in the neutralino sector. Hence, we 
consider in this work both the decays and calculate in each case the top 
polarization as a function of model parameters.

This paper is organized as follows. Section~\ref{sec:2} discusses the 
evaluation of top polarization in the rest frames of a gluino and a stop 
respectively. Section~\ref{sec:3} describes the formalism of our work and a 
derivation of our main result. Section~\ref{sec:3a} describes the procedure to 
obtain the top polarization at the level of $pp$ collisions. In Section \ref{sec:4} we describe the numerical work with which we validate our analytical result. We conclude in Section~\ref{sec:5} and present some of the calculational details in the appendix.
  
\section{Top polarization in the rest frame of the decaying particle}
\label{sec:2}
\subsection{Gluino decay}
The gluino decay mode of interest are the ones involving the top quark: 
$\tilde{g}\rightarrow t\tilde{t}_1^{\ast}$ and 
$\tilde{g}\rightarrow t\tilde{t}_2^{\ast}$. Here $\tilde{t}_1$ and 
$\tilde{t}_2$ are the lighter and the heavier of stop mass eigenstates, 
respectively. The interaction of a top, a gluino and a stop mass eigenstate depends only on the stop 
mixing angle $\theta_{\tilde{t}}$, when mixing with first two generations is neglected. The stop mixing angle relates the two mass 
eigenstates of the stop ($\tilde{t}_1$, $\tilde{t}_2$) to their interaction 
eigenstates $\tilde{t}_L$ and $\tilde{t}_R$: 
\begin{align}
\tilde{t}_1&=\cos\theta_{\tilde{t}}\tilde{t}_L+\sin\theta_{\tilde{t}}
\tilde{t}_R,\\\nonumber
\tilde{t}_2&=-\sin\theta_{\tilde{t}}\tilde{t}_L+\cos\theta_{\tilde{t}}
\tilde{t}_R.
\end{align}
The interaction of the stop with the gluino and the top is given by the 
following Lagrangian, again in the approximation that there is no flavor-mixing 
between the first two generations and the third generation:
\begin{equation}
\mathcal{L}_{\tilde{t}\tilde{g}t}=-\sqrt{2}g_3\left\{\bar{t}
[W_{3i}P_R-W_{6i}P_L]T^{a}\tilde{g}^{a}\tilde{t}_i\right\}
+\mathrm{h}.\mathrm{c}.
\end{equation}
In the above expression, $g_3$ denotes the strong coupling constant, $i=3,6$
the stop mass eigenstates ($\tilde{t}_1,\tilde{t}_2$), $a=1,\cdots 8$ the 
adjoint $SU(3)_c$ indices and $P_L$, $P_R$ the chirality projecting operators.
The color indices of the top and the stop mass eigenstates have been 
suppressed. The sfermion mixing matrix is denoted by $W$ and it's elements are:
\begin{equation}\label{eq:stopmix}
  W_{33}=W_{66}=\cos\theta_{\tilde{t}}, \hspace{0.5cm}
  W_{36}=-W_{63}=\sin\theta_{\tilde{t}}
\end{equation}
where $\theta_{\tilde{t}}$ is the stop mixing angle.
To derive the expression for top polarization in the gluino rest frame, one 
begins by writing down the amplitude for the process and evaluating the partial 
width $\Gamma^{\lambda}$ of the gluino decaying into a stop and a top with a 
helicity $\lambda$. Then the top polarization in the gluino rest frame is given
by the formula:
\begin{equation}\label{eq:4}
\mathcal{P}_0=\frac{\Gamma^{+}-\Gamma^{-}}{\Gamma^{+}+\Gamma^{-}}.
\end{equation}
The amplitudes for the gluino decay $\tilde{g}\rightarrow t\tilde{t}_1^{\ast}$ 
is as follows:
\begin{eqnarray}
\mathcal{M}(\tilde{g}\rightarrow \tilde{t}_1t)&=&-\sqrt{2} \ g_3 \ T^{a}  \\
&\times& \bar{u}(p_t,\lambda_t)  (\cos\theta_{\tilde{t}}P_R-\sin\theta_{\tilde{t}}
P_L) \ u(p_{\tilde{g}},\lambda_{\tilde{g}})\nonumber
\end{eqnarray} 
where $\lambda_t$ and $\lambda_{\tilde{g}}$ denote the helicities of the top, 
and the gluino, respectively and $t^{a}$ is the color factor.  Squaring the amplitude and taking average over 
helicities and color indices of initial state particles and summing over the 
color indices of the top and the stop, we get
\begin{eqnarray}
\Gamma^{\lambda}&\propto& \frac{g_3^2}{6}\left[(m_{\tilde{g}}^2-
m_{\tilde{t}}^2+m_t^2-2m_{\tilde{g}}m_t\sin 2\theta_{\tilde{t}})\right.\nonumber\\
&&\left.-2\lambda 
\cos 2\theta_{\tilde{t}} \ K^{1/2}(m_{\tilde{g}}^2,m_{\tilde{t}}^2,m_t^2)\right].
\end{eqnarray}
In the above expression, $K(x,y,z)=x^2+y^2+z^2-2 x y-2 y z -2zx$.
 Introducing the notation $\xi_t=m_{t}^2/m_{\tilde{g}}^2$ and 
$\xi_{\tilde{t}_1}=m_{\tilde{t}_1}^2/m_{\tilde{g}}^2$, we get
\begin{equation}\label{eq:5}
\mathcal{P}_0=\frac{-K^{1/2}(1,\xi_t,\xi_{\tilde{t}_1})\cos 2\theta_{\tilde{t}}}
{1-\xi_{\tilde{t}_1}+\xi_t-2\sqrt{\xi_t}\sin 2\theta_{\tilde{t}}}.
\end{equation}
This choice of using top helicity states for spin states corresponds to the 
so-called helicity basis for the top polarization. We use this basis throughout
this work. 

\subsection{Stop decay}
The stop can decay in a number of modes, e.g., $\tilde{t}_1\rightarrow 
\tilde{\chi}_i^0t$, and $\tilde{t}_1\rightarrow \tilde{\chi}_i^{+}b$ etc. We 
consider the decays where a top quark is produced: $\tilde{t}_1\rightarrow t
\tilde{\chi}_i^0$ and $\tilde{t}_2\rightarrow t\tilde{\chi}_i^0$ ($i=1,\cdots,
4$). The vertex corresponding to the decay of a stop mass eigenstate, 
$\tilde{t}_{1,2}\rightarrow t\tilde{\chi}_1^0$, is given by the following 
Lagrangian,
\begin{align}
\mathcal{L}_{\tilde{t}\tilde{\chi}^0t}&=\overline{\tilde{\chi}_1^0}
\left[G^LP_L+G^RP_R\right]\tilde{t}_1^{\dagger}t +\mathrm{h}.\mathrm{c}.
\end{align}
The $G^L$ and $G^R$ in the above equation are as follows: 
\begin{eqnarray}\label{eq:6}
G^L&=&-\frac{g_2}{\sqrt{2}}(N_{12}+\frac{1}{3}\tan\theta_W N_{11})
\cos\theta_{\tilde{t}}-Y_tN_{14}\sin\theta_{\tilde{t}}\nonumber\\
G^R&=&-Y_tN_{14}^{\ast}\cos\theta_{\tilde{t}}+g_2\left(\frac{2\sqrt{2}}{3}
\tan\theta_WN_{11}\sin\theta_{\tilde{t}}\right)
\end{eqnarray}  
where $g_2$ and $Y_t$ are the $SU(2)_L$ and top Yukawa couplings, respectively 
and $\theta_W$ is the Weak mixing angle. $N_{ij}, \ (i,j=1,\cdots,4)$ in the 
above equation are the elements of the $4\times 4$ neutralino mixing matrix.
Note that the top quark is in the final state. This means that $G^L$ and $G^R$ 
correspond to the couplings of left and right chiral top quarks, respectively. 
The neutralino mixing matrix is the diagonalizing matrix of the mass matrix of 
the neutral gauginos (the bino, the wino) and the neutral higgsinos.

The expression for polarization of top quark produced in the process 
$\tilde{t}_1\rightarrow t\tilde{\chi}_1^0$,  evaluated in the stop rest frame 
is given in~\cite{Nojiri:1994it,Kitano:2002ss,Perelstein:2008zt,
Belanger:2012tm,Shelton:2008nq}. We sketch the derivation of expression for 
top polarization in the stop rest frame for sake of completeness. The amplitude 
for the stop decay $\tilde{t}_1\rightarrow t\tilde{\chi}_1^0$ is given by:
\begin{equation}
\mathcal{M}(\tilde{t}_1\rightarrow t\tilde{\chi}_1^0)=\bar{u}(p_t,\lambda_t)
\left[G^LP_R+G^RP_L\right]v(p_{\tilde{\chi}},\lambda_{\tilde{\chi}}).
\end{equation}
Computing $\Gamma^{\pm}$, the partial width of the stop decaying into a 
neutralino and a top with a helicity $\lambda=\pm$ and using the expression 
Eq.~(\ref{eq:4}), we get the required expression for the top polarization in stop 
rest frame as: 
\begin{equation}\label{eq:7}
\mathcal{P}_0=\frac{(|G^R|^2-|G^L|^2) \ K^{1/2}(1,\eta_t,\eta_{\tilde{\chi}})}
{(|G^R|^2+|G^L|^2)(1-\eta_t-\eta_{\tilde{\chi}})
-4\sqrt{\eta_t\eta_{\tilde{\chi}}}\operatorname{Re}(G^LG^{R\ast})}
\end{equation}
where $\eta_t=m_t^2/m_{\tilde{t}_1}^2$ and $\eta_{\tilde{\chi}}=
m_{\tilde{\chi}_1^0}^2/m_{\tilde{t}_1}^2$.

\begin{table}
\begin{ruledtabular}
\begin{tabular}{cccc}
parameter & BP1   &  BP2    &   BP3 \\
\hline
$m_{\tilde{g}}$ & 2290 & 2291 & 608\\
$m_{\tilde{\chi}_1^0}$ & 248 & 248 & 97\\
$m_{\tilde{t}_1}$ & 498 & 493 & 400\\
$\Gamma_{\tilde{g}}$ & 235 & 203 & 5.5\\
$\Gamma_{\tilde{t}_1}$ & 2.4 & 6.0 & 2.0\\
\hline
$\sin\theta_{\tilde{t}}$ & 0.0644 & 0.9979 & 0.8327\\
$\cos\theta_{\tilde{t}}$ & 0.9979 & 0.0642 & 0.5536\\
$N_{11}$ & 0.0953 & -0.0988 & 0.9863\\
$N_{12}$ & -0.0637 & 0.0619 & -0.0531\\
$N_{14}$ & -0.6939 & 0.6937 & -0.0531\\
$y_t$ & 0.8507 & 0.8508 & 0.8928\\
\end{tabular}
\end{ruledtabular}
\caption[The benchmark points]{The benchmark points used in this work. 
Masses and widths of the particles are given in GeV.}
\label{tab:1}
\end{table}
\subsection{Polarization and mixing angles}
Discussions of using  polarization of the top quark produced in stop/sbottom 
decays $\tilde{t}_1\rightarrow t\tilde{\chi}_1^0$ and $\tilde{b}_i\rightarrow 
t\tilde{\chi}_j^-$ ($i,j=1,2$) to probe the mixing angle in the stop/sbottom 
sector exist in literature (see, for example, Refs.~\cite{Perelstein:2008zt,
Belanger:2013gha,Boos:2003vf,Gajdosik:2004ed}). But, as one can easily see from
Eq.~(\ref{eq:5}), top polarization in the gluino decays can probe the mixing angle
in the stop sector irrespective of the mixing in the neutralino sector, given a
value of $\Delta m_{\tilde g}=m_{\tilde{g}}-m_{\tilde{t}_1}$. This is possible 
only when there is a large mass difference between $\tilde{t}_1$ and 
$\tilde{t}_2$ . This can be seen as follows. Equation (\ref{eq:5}) gives the 
expression for the top polarization that is produced along with $\tilde{t}_1$ 
in the gluino decay. The corresponding expression for top polarization in the 
case of $\tilde{g}\rightarrow \tilde{t}_2t$ can be obtained by the following 
interchange: $\sin\theta_{\tilde{t}}\rightarrow \cos\theta_{\tilde{t}}$,  
$\cos\theta_{\tilde{t}} \rightarrow -\sin\theta_{\tilde{t}}$ and 
$m_{\tilde{t}_1}\rightarrow m_{\tilde{t}_2}$. This means that 
$\cos 2\theta_{\tilde{t}}\rightarrow -\cos 2\theta_{\tilde{t}}$ and 
$\sin 2\theta_{\tilde{t}}\rightarrow -\sin 2\theta_{\tilde{t}}$ which changes 
the sign of the top polarization. If we count the tops from both 
$\tilde{g}\rightarrow \tilde{t}_1t$ and $\tilde{g}\rightarrow \tilde{t}_2t$ and
if both stops are degenerate we get unpolarized tops. But, if there is a large
mass difference between the two stops the net top polarization can be nonzero.
When models in MSSM with a natural Higgs sector are realized in nature, we 
expect a large mass difference between the two stop mass eigenstates, as 
mentioned before. If we assume that the heavy stop $\tilde{t}_2$ is heavier 
than the gluino the tops can be polarized and this is the scenario  which we 
focus on in this paper. 
In the case of stop decay, we consider decays of only light stop mass 
eigenstate $\tilde{t}_1$ since it can be accessible at colliders and partly 
because we focus on scenarios where the Higgs sector is natural. Hence, we study polarization of the top produced in the decay of a gluino into a top and $\tilde{t}_1$, for a few benchmark points.
We also assume that  the neutralino produced in the stop decay is the 
lightest of the four neutralino states.
\begin{table}
\begin{ruledtabular}
\begin{tabular}{cccc}
Benchmark $\to$ & BP1 & BP2 & BP3  \\ \hline
$\mathcal{P}_0$     & $-1.00$ &  $+1.00$ &  $+0.98$ \\
$\mathcal{P}_{MC}$  & $-0.99$ &  $+0.99$ &  $+0.51$ \\
$\mathcal{P}_{NW}$ & $-1.00$ &  $+1.00$ &  $+0.50$ \\
$\mathcal{P}_{BW}$  & $-0.98$ &  $+0.98$ &  $+0.51$ \\
\end{tabular}
\end{ruledtabular}
\caption{List of top polarization in gluino decay at $\sqrt{s}=13$ TeV LHC 
calculated for three benchmark in gluino rest frame $\mathcal{P}_0$, 
in the lab frame $\mathcal{P}_{MC}$ and using estimators 
$\mathcal{P}_{NW}$ and $\mathcal{P}_{BW}$ (see sections~\ref{sec:5} and 
\ref{sec:4}). 
The benchmark points BP1, BP2, and BP3 are given in Table~\ref{tab:1}.}
\label{tab:2}
\end{table}
\begin{table}
\begin{ruledtabular}
\begin{tabular}{cccc}
Benchmark $\to$ & BP1 & BP2 & BP3  \\ \hline
$\mathcal{P}_0$     & $+0.92$ &  $-0.93$ &  $+0.97$ \\
$\mathcal{P}_{MC}$  & $+0.61$ &  $-0.60$ &  $+0.65$ \\
$\mathcal{P}_{NW}$ & $+0.61$ &  $-0.60$ &  $+0.65$ \\
$\mathcal{P}_{BW}$  & $+0.61$ &  $-0.60$ &  $+0.65$ \\
\end{tabular}
\end{ruledtabular}
\caption{List of top polarization in stop decay at $\sqrt{s}=13$ TeV LHC. Rest
of the details are same as in Table~\ref{tab:2}.}
\label{tab:3}
\end{table}
The bench mark points used for numerical simulations are listed in 
Table~\ref{tab:1} and corresponding top polarizations are listed in 
Tables~\ref{tab:2} and \ref{tab:3} for reference. Here we have listed the rest
frame polarization $\mathcal{P}_0$ along with the lab frame value 
$\mathcal{P}_{MC}$ obtained using full Monte-Carlo simulations. The lab frame 
values are usually reduced in magnitude due to change of quantization basis, as
will be discussed in the next section. The proposed polarization estimators 
with narrow-width-approximation ($\mathcal{P}_{NW}$) and with Breit-Wigner
folding ($\mathcal{P}_{BW}$), which can be directly compared to $\mathcal{P}_{MC}$, are also listed in Tables~\ref{tab:2} and
\ref{tab:3} for comparisons and will be discussed in latter sections.

\section{General formalism}\label{sec:3}
The cross section for the process in Eq.~(\ref{eq:1}) can be written as
\begin{eqnarray}
\sigma&=&\sum_{q_1,q_2}\int dx_1dx_2 \ f_{q_1/p}(x_1)f_{q_2/p}(x_2) \nonumber\\
&&\times \ \ \hat{\sigma}(q_1q_2\rightarrow AA\rightarrow A+t+B)
\end{eqnarray}
where the sum extends over all the parton flavors, $q_1$ and $q_2$. 
%
Since the top quark also decays, we can access its polarization through angular
distribution of its decay products in the rest frame of the top quark. In the 
semi-leptonic decay of the top quark, the direction of motion of the charged 
lepton is $100\%$ correlated with the top polarization, at 
leading order. In the top rest frame, we have:
\begin{equation}\label{eq:8}
\frac{1}{\Gamma}\frac{d\Gamma}{d\cos\theta_{\ell}}=\frac{1}{2}
(1+\mathcal{P}_0 \ \alpha_{\ell} \ \cos\theta_{\ell}).
\end{equation}
The coefficient $\alpha$ in the above equation is called the spin analyzing 
power and it is maximal for the charged lepton ($\alpha_{\ell}=1$ at the LO of 
the SM). The value of $\mathcal{P}_0$ depends on the choice of quantization 
axis of the  top quark. 
When the top spin quantization axis taken as its direction of motion in the 
rest frame of the gluino or the stop, the value of $\mathcal{P}_0$ is given 
by Eqs. (\ref{eq:5}) and (\ref{eq:7}) respectively.  

The full amplitude of the process under consideration is given by the following
expression:
\begin{eqnarray}\label{eq:9}
\mathcal{M}&\sim& \mathcal{M}'(p_1p_2\rightarrow AA)_{\alpha}\left(
\frac{\slashed{p}_A+m_A}{(p_A^2-m_A^2)+im_A\Gamma_A}\right)_{\alpha'\alpha}
\nonumber\\ &\times&
\mathcal{M}'(A\rightarrow tB)_{\beta\alpha'}
\left(\frac{\slashed{p}_t+m_t}{(p_t^2-m_t^2)+im_t\Gamma_t}
\right)_{\beta'\beta}\nonumber\\
&\times& \mathcal{M}'(t\rightarrow b\bar{\ell}\nu)_{\beta'}
\end{eqnarray} 
where the prime indicates that the amplitudes do not have their external 
fermion wave functions which are part of the propagators, the Greek indices 
denote the components of Dirac matrices and repeated indices are summed over. 
Squaring the amplitude and summing/averaging over the spins and color indices 
of the external states (which are suppressed in these expressions) gives the 
propagator factors of the form
\begin{equation}\label{eq:10}
\frac{1}{(p_A^2-m_A^2)^2+m_A^2\Gamma_A^2}.
\end{equation} 
When the width of a particle is much smaller than its mass, the narrow width 
approximation (NWA) can be used which consists of the following replacement for
the propagators:
\begin{equation}\label{eq:11}
\frac{1}{(p^2-m_A^2)^2+\Gamma_A^2m_A^2}\rightarrow \frac{\pi}{m_A\Gamma_a}
\delta(p^2-m_A^2)\theta(p^0)
\end{equation}
Similar replacements can be made for all the other intermediate particles, 
including the top quark.
Under this approximation, the production of a particle and its decay are separated as factors with spin-correlations. This is made possible, in the narrow width approximation, because of the following relation for the numerator of a propagator (for a spin-1/2 particle)
\begin{equation}\label{eq:12}
\slashed{p}+m=\sum_{\lambda}u(p,\lambda)\bar{u}(p,\lambda)
\end{equation}
where $\lambda$ denotes the helicity state of the particle, defined with 
respect to the momentum $p$ (its spin quantization axis). Using this relation 
in the numerator of Eq.~(\ref{eq:9}), we get,
\begin{eqnarray}\label{eq:13}
\mathcal{M}^{\mathrm{num}}&\sim& \sum_{\lambda_A,\lambda_t}
\bar{u}(p_A,\lambda_A)\mathcal{M}^{\prime \mathrm{num}}(p_1p_2\rightarrow AA)
\nonumber \\
&&\times \ \ \bar{u}(p_t,\lambda_t)
\mathcal{M}^{\prime \mathrm{num}}(A\rightarrow tB)u(p_A,\lambda_A) \nonumber\\
&&\times \ \ 
\mathcal{M}^{\prime \mathrm{num}}(t\rightarrow b\bar{\ell}\nu) u(p_t,\lambda_t)
\\\nonumber
&=&\sum_{\lambda_A,\lambda_t}
\mathcal{M}^{\mathrm{num}}(p_1p_2\rightarrow AA)_{\lambda_A} 
\mathcal{M}^{\mathrm{num}}(A\rightarrow tB)_{\lambda_A,\lambda_t}\nonumber\\
&&\times \mathcal{M}^{\mathrm{num}}(t\rightarrow b\bar{\ell}\nu)_{\lambda_t}
\end{eqnarray}
Squaring the amplitude $\mathcal{M}$ and performing the replacements of 
Eq.~(\ref{eq:11}), in the case of gluino ($A=\tilde{g}$), we get
\begin{equation}\label{eq:14}
\hat{\sigma}(\hat{s})=\int \hspace{-0.1cm}d\Omega
\sum_{\{\lambda\}}
\left(\frac{d\hat{\sigma}_{\tilde{g}\tilde{g}}}{d\Omega}
\right)_{\lambda_{\tilde{g}}\lambda_{\tilde{g}}'}\left(\frac{d
\Gamma_{\tilde{g}}}{\Gamma_{\tilde{g}}}\right)^{\lambda_{\tilde{g}}
\lambda_{\tilde{g}}'}_{\lambda_t\lambda_t'}\left(\frac{d\Gamma_t}{\Gamma_t}
\right)_{\lambda_t\lambda_t'}
\end{equation}
at the parton level. Similar parton level calculations in the case of stop 
yields a simpler expression, given as:
\begin{equation}\label{eq:15}
\hat{\sigma}(\hat{s})=\int d\Omega\sum_{\{\lambda\}}
\frac{d\hat{\sigma}_{\tilde{t}\tilde{t}}}{d\Omega}
\left(\frac{d\Gamma_{\tilde{t}}}{\Gamma_{\tilde{t}}}\right)_{
\lambda_t\lambda_t'}
\left(\frac{d\Gamma_t}{\Gamma_t}\right)_{\lambda_{t}\lambda_t'}.
\end{equation}
In these expressions, the parton level differential cross section 
(density matrix) for the pair production of gluinos is denoted by $\left(d\hat{\sigma}_{\tilde{g}\tilde{g}}/d\Omega\right)_{\lambda_{\tilde{g}}\lambda_{\tilde{g}}'}$. The spin indices (helicities, in our case) of all other intermediate particles are summed over. 

When the gluino is unpolarized, as it would be in the case of QCD production, 
the production cross section (density matrix) for gluino pairs (see 
Eq.~(\ref{eq:14})) can be written as 
$(d\hat{\sigma}_{\tilde{g}\tilde{g}}/d\Omega)_{\lambda_{\tilde{g}},
\lambda_{\tilde{g}}^{\prime}}=(d\hat{\sigma}_{\tilde{g}\tilde{g}}/d\Omega) \ 
(\delta^{\lambda_{\tilde{g}}\lambda_{\tilde{g}}^{\prime}}/2)$ where 
$\lambda_{\tilde{g}}, \ \lambda_{\tilde{g}}^{\prime}$ are the helicity states 
of the intermediate gluino. On summing over gluino helicities, the differential
cross-section of gluino pair production becomes a multiplicative factor in just
the same way as $d\hat{\sigma}_{\tilde{t}\tilde{t}}/d\Omega$ does in the case 
of stop decay (Eq.~(\ref{eq:15})).  We emphasize here that the helicities are 
defined in the frame in which the top polarization needs to be defined. We 
first take the frame in which Eq.~(\ref{eq:14}) and Eq.~(\ref{eq:15}) are defined to
be the corresponding parton center of mass (PCM) frame. This frame can be 
reached from the top rest frame in two ways: (i) a direct Lorentz boost along the
direction of top in the PCM frame, (ii) a Lorentz boost to the rest frame of the 
mother particle along the direction of top momentum in that frame followed by a 
Lorentz transformation to the PCM frame which is, in general, not in the 
direction of top momentum. As result, the helicity states of the top in the two
cases are not identical. This affects the value of top polarization measured in
the PCM frame. It turns out that the helicity states of the top in the latter 
case can be written as a rotation of the helicity states defined in the PCM 
frame. Note that this procedure is also applicable to the calculation of top 
polarization in the lab frame. However, the top polarization in lab frame can 
be obtained without going through the parton center of mass frame, as will be 
discussed in Section \ref{sec:3a}. 
     
\subsection{Helicity and Lorentz boosts}\label{ssec:1}
As mentioned before, the helicity of a particle is not invariant under 
Lorentz boosts in general. The helicity state $|p,\lambda\rangle $ transforms 
under a Lorentz boost as
\begin{equation}\label{eq:16}
|p,\lambda\rangle=R_{\lambda
\lambda^{\prime}}|p^{\prime},\lambda^{\prime}\rangle
\end{equation}
where $|p^{\prime},\lambda^{\prime}\rangle$ is the helicity state of the 
particle after the Lorentz boost has been applied \cite{Bourrely:1980mr}. The 
helicity states $|p,\lambda\rangle$ are constructed from the eigen states of 
spin in the rest frame of the particle by a series of transformations: 
$|p,\lambda\rangle =R_z(\phi)R_y(\theta)\Lambda_z(\beta)|m,s_z=\lambda\rangle$,
where $\theta,\phi$ and $\beta$ are the angles and the velocity of the 
particle. With this convention for helicity states,  the coefficients 
$R_{\lambda,\lambda^{\prime}}$ in Eq.~(\ref{eq:16}) can be given in the 
following form of a rotation matrix 
\begin{equation}\label{eq:17}
R=R_z(-\chi)R_y(-\omega)
\end{equation}
where, $\chi$ and $\omega$ are some angles which depend on the direction and 
magnitude of the Lorentz boost applied on $|p,\lambda\rangle$. Expressions for
$\chi$ and $\omega$ are given in Appendix~\ref{sec:app}.
\subsection{Gluino decay}\label{ssec:2}
In the rest frame of the gluino\footnote{We have taken $z$-axis of the lab 
frame coordinate system along one of the beam directions and the $x$-axis in a 
plane containing the top momentum and the beam axis and the $y$- axis along the
normal to this plane. The azimuthal angle of the top, in the parton center of 
mass frame and in the lab frame $\phi=0$, due to this choice of the coordinate
system. All the angles that have been mentioned so far correspond to this 
coordinate system.} which decays into a top, the differential decay width which
appears in Eq.~(\ref{eq:14}) is given by
\begin{eqnarray}\label{eq:18}
\int\!\! \left(\frac{\mathrm{d}\Gamma_{\tilde g}}{\Gamma_{\tilde g}}\right)_{
\lambda\lambda^{\prime}}
&=&\frac{m_g}{64\pi^2\Gamma_{\tilde g}} \ \frac{4\pi\alpha_S}{9} \int\!\! 
\mathrm{d}\cos\theta_t^{\tilde{g}} \ \mathrm{d}\phi_t^{\tilde{g}}
 \nonumber\\
&\times& \Big[\delta_{\lambda\lambda^{\prime}}\left(\frac{1}{2}
(1-\xi_{\tilde{t}}+\xi_t)-\sqrt{\xi_t}\sin 2\theta_{\tilde{t}}\right)
\nonumber\\
&&-\cos 2\theta_{\tilde{t}}\sqrt{\xi_t}\beta\gamma
\sigma^{3}_{\lambda\lambda^{\prime}}\Big]. 
\end{eqnarray}
$\xi_{t},\xi_{\tilde{t}_1}$ have been defined in Section \ref{sec:2}. The 
velocity of the top in the above equation is given by 
$\beta=K^{1/2}(1,\xi_t,\xi_{\tilde{t}_1})/(1-\xi_{\tilde{t}_1}+\xi_t)$ 
and $\gamma=1/\sqrt{1-\beta^2}$. The helicity states of the top quark 
$\lambda,\lambda^{\prime}$ in the above equation are defined with respect to 
the top momentum in the gluino rest frame. For top decaying in the 
semi-leptonic channel $t\rightarrow b\bar{\ell}\nu$ (with narrow width 
approximation for the top as well as the $W$ boson) the differential 
decay width is given by
\begin{align}\label{eq:19}
\int \!\! \frac{\mathrm{d}\Gamma_t}{\Gamma_t}&=
\int \!\! \mathrm{d}x_{\ell}\mathrm{d}\Omega_{\ell} \ 
\frac{G_F^2m_t^5 \ x_{\ell}(1-x_{\ell})}{64\pi^3\bar{\Gamma}_Wr_t\Gamma_t} \ 
(\delta_{\lambda\lambda^{\prime}}+\hat{p}_{\ell}
\cdot\vec{\sigma}_{\lambda\lambda^{\prime}})
\end{align}
where $r_t=m_t^2/m_W^2$, $\bar{\Gamma}_W=\Gamma_W/m_W$, $x_{\ell}=2E_{\ell}/m_t$ with $1/r_t\leq x_{\ell}\leq 1$. The direction of the momentum of the charged lepton from decay of the top, in the top rest frame, is denoted in the above equation as $\hat{p}_{\ell}$.
Hence,
\begin{equation}\label{eq:20}
\frac{d\Gamma_{\tilde{g}}}{\Gamma_{\tilde{g}}}^{\mathrm{PCM}}=R\left(\frac{d\Gamma_{\tilde{g}}}{\Gamma_{\tilde{g}}}\right)R^{\dagger},
\end{equation}
and 
\begin{equation}\label{eq:21}
\frac{d\Gamma_t}{\Gamma_t}^{\mathrm{PCM}}=\frac{d\Gamma_t}{\Gamma_t}.
\end{equation}
The rotation matrix $R$ in Eq.~(\ref{eq:20}) is  given in Eq.~(\ref{eq:17}). 
Substituting the expressions for the differential decay widths for 
$\tilde{g}\rightarrow t\tilde{t}_1^{\ast}$ and $t\rightarrow b\bar{\ell}\nu$ 
from Eqs.~(\ref{eq:18}), (\ref{eq:19}), we get
\begin{align}\label{eq:22}
\frac{\mathrm{d}\hat{\sigma}}{\mathrm{d}x_{\ell}\mathrm{d}\cos\theta_{t\ell}
\mathrm{d}\phi_{t\ell}}&\propto\int\!\!
\frac{\mathrm{d}\hat{\sigma}_{\tilde{g}\tilde{g}}}
{\mathrm{d}\Omega_{\tilde{g}}} \ 
\Big[1+\mathcal{P}_0 \ \cos\omega\cos\theta_{t\ell}\\\nonumber
&-\mathcal{P}_0 \ \sin\omega\sin\theta_{t\ell}\cos(\phi_{t\ell}+\chi)\Big]
\\\nonumber
&\times x_{\ell}(1-x_{\ell}) \ \mathrm{d}\Omega_g \ 
\mathrm{d}\Omega_t^{\tilde{g}}.
\end{align}
where $\hat{\sigma}$ is the cross section for the full parton level process 
$q_1q_2\rightarrow \tilde{g}\tilde{g}\rightarrow \tilde{g}\tilde{t}_1^{\ast}
b\bar{\ell}\nu$, $\mathcal{P}_0$ is the value of top polarization in the rest 
frame of the gluino.

Let $I(\bar{\beta}(\hat{s}))=\frac{1}{2}\int \cos\omega \,\,
\mathrm{d}\cos\theta_t^{\tilde{g}}$. Integrating Eq.~(\ref{eq:22}) over all 
variables except $\theta_{t\ell}$, we get
\begin{eqnarray}\label{eq:24}
\frac{\mathrm{d}\hat{\sigma}}{\mathrm{d}\cos\theta_{t\ell}}&=&
\hat{\sigma}_{\tilde{g}\tilde{g}}(\hat{s}) \ 
\mathcal{B}(\tilde{g}\rightarrow t\tilde{t}_1^{\ast}) \ 
\mathcal{B}(t\rightarrow b\bar{\ell}\nu)\nonumber\\
&&\times \frac{1}{2}(1+\mathcal{P}_0 I(\bar{\beta}) \ \cos\theta_{t\ell})
\end{eqnarray}
where $\hat{\sigma}_{\tilde{g}\tilde{g}}(\hat{s})$ denotes the parton level 
cross-section for pair production of gluinos. Note that the angle $\chi$ drops 
out of Eq.~(\ref{eq:24}). The coefficient of $\cos\theta_{t\ell}$ can be 
interpreted as the polarization of the top as given by the lepton angular 
distribution in the top rest frame. 
The helicity rotation angle $\omega$ is independent of the direction of motion 
of the gluino in the parton center of mass frame (see Appendix \ref{sec:app}), 
so is the polarization of the top.
The expressions for $I(\bar{\beta})$ are as follows:
\begin{eqnarray}\label{eq:25}
I(\bar{\beta})&=&\frac{1}{2\beta^2\bar{\beta}}\left[2\bar{\beta}-(1-\beta^2)\log\left(\frac{1+\bar{\beta}}{1-\bar{\beta}}\right)\right]\,\,\,\,(\beta>\bar{\beta})\nonumber\\
&=&\frac{1}{2\beta^2\bar{\beta}}\left[2\beta-(1-\beta^2)\log\left(\frac{1+\beta}{1-\beta}\right)\right]\,\,\,\,(\beta < \bar{\beta})\nonumber\\
\end{eqnarray} 
Substituting the value of top velocity $\beta$ in the gluino rest frame,
we get the following expression which is valid for both cases, viz., 
$\beta>\bar{\beta}$ and $\beta<\bar{\beta}$:
\begin{eqnarray}\label{eq:27}
\mathcal{P}(\bar{\beta})&=&I(\bar{\beta})\mathcal{P}_0\nonumber\\
&=&\frac{\mathcal{P}_0}{2\bar{\beta}K}\times\Big[(1-\xi_{\tilde{t}_1}+\xi_t)\Big((K^{1/2}+\bar{\beta}(1-\xi_{\tilde{t}_1}+\xi_t))\nonumber\\
&&-|K^{1/2}-\bar{\beta}(1-\xi_{\tilde{t}_1}+\xi_t)|\Big)\nonumber\\
&&-4\xi_t\log\left(\frac{\Delta_1+K^{1/2}+\bar{\beta}(1-\xi_{\tilde{t}_1}+\xi_t)}{\Delta_2+|K^{1/2}-\bar{\beta}(1-\xi_{\tilde{t}_1}+\xi_t)|}\right)\Big]
\end{eqnarray}
where, $K\equiv K(1,\xi_{\tilde{t}_1},\xi_t)$ and $\Delta_{1,2}=(1-\xi_{\tilde{t}_1}+\xi_t)(1\pm \beta\bar{\beta})=(1-\xi_{\tilde{t}_1}+\xi_t)\pm\bar{\beta}K^{1/2}(1,\xi_{\tilde{t}_1},\xi_t)$. This expression can also be written as
\begin{eqnarray}\label{eq:28}
\mathcal{P}(\bar{\beta})&=&\frac{\mathcal{P}_0}{2\bar{\beta}K}
\Bigg[(1-\xi_{\tilde{t}_1}+\xi_t)\nonumber\\
&&\times\Big(\sqrt{\Delta_1^2-4(1-\bar{\beta}^2)\xi_t}-\sqrt{\Delta_2^2-4(1-\bar{\beta}^2)\xi_t}\Big)\nonumber\\
&-&4\xi_t\log\left(\frac{\Delta_1+\sqrt{\Delta_1^2-4(1-\bar{\beta}^2)\xi_t}}{\Delta_2+\sqrt{\Delta_2^2-4(1-\bar{\beta}^2)\xi_t}}\right)\Bigg].
\end{eqnarray}

\subsection{Stop decay}\label{ssec:3}
Analogous to Eq.~(\ref{eq:28}), the expression for polarization of the top, in 
the PCM frame where the mother stop is moving with a velocity $\bar{\beta}$ is 
the following:
\begin{eqnarray}\label{eq:30}
\mathcal{P}(\bar{\beta})&=&\frac{\mathcal{P}_0}{2\bar{\beta}K}
\Bigg[(1+\eta_t-\eta_{\tilde{\chi}})\nonumber\\
&&\times\left(\sqrt{\Delta_1^2-4
(1-\bar{\beta}^2)\eta_t}-\sqrt{\Delta_2^2-4(1-\bar{\beta}^2)\eta_t}\right)
\nonumber\\
&-&4\eta_t\log\left(\frac{\Delta_1+\sqrt{\Delta_1^2-4(1-\bar{\beta}^2)\eta_t}}
{\Delta_2+\sqrt{\Delta_2^2-4(1-\bar{\beta}^2)\eta_t}}\right)\Bigg]
\end{eqnarray}
 where $\Delta_{1,2}=1+\eta_t-\eta_{\tilde{\chi}}\pm \bar{\beta}
K^{1/2}(1,\eta_t,\eta_{\tilde{\chi}})$.
\begin{figure}
\includegraphics[scale=0.45,keepaspectratio=true]{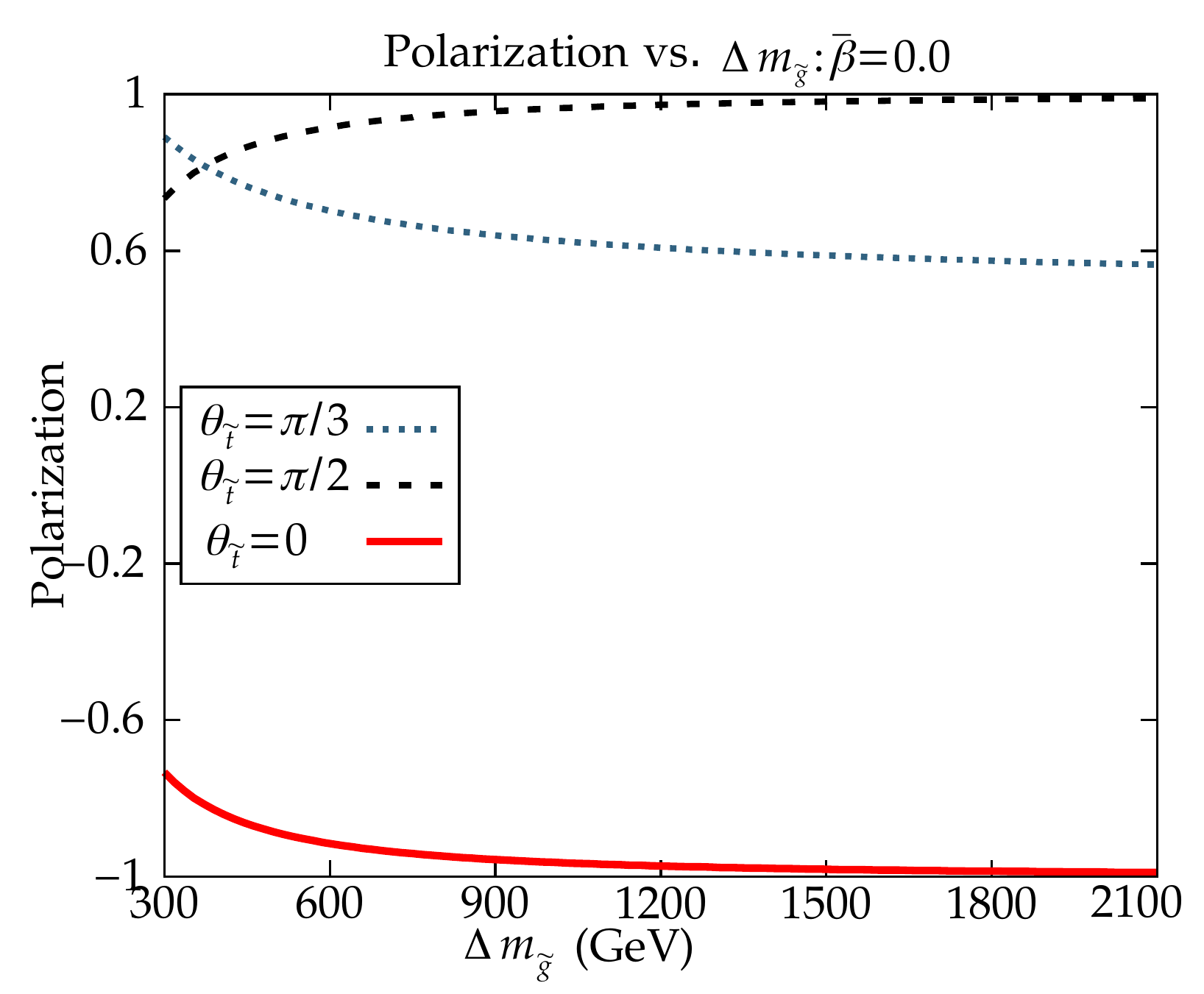}
\includegraphics[scale=0.45,keepaspectratio=true]{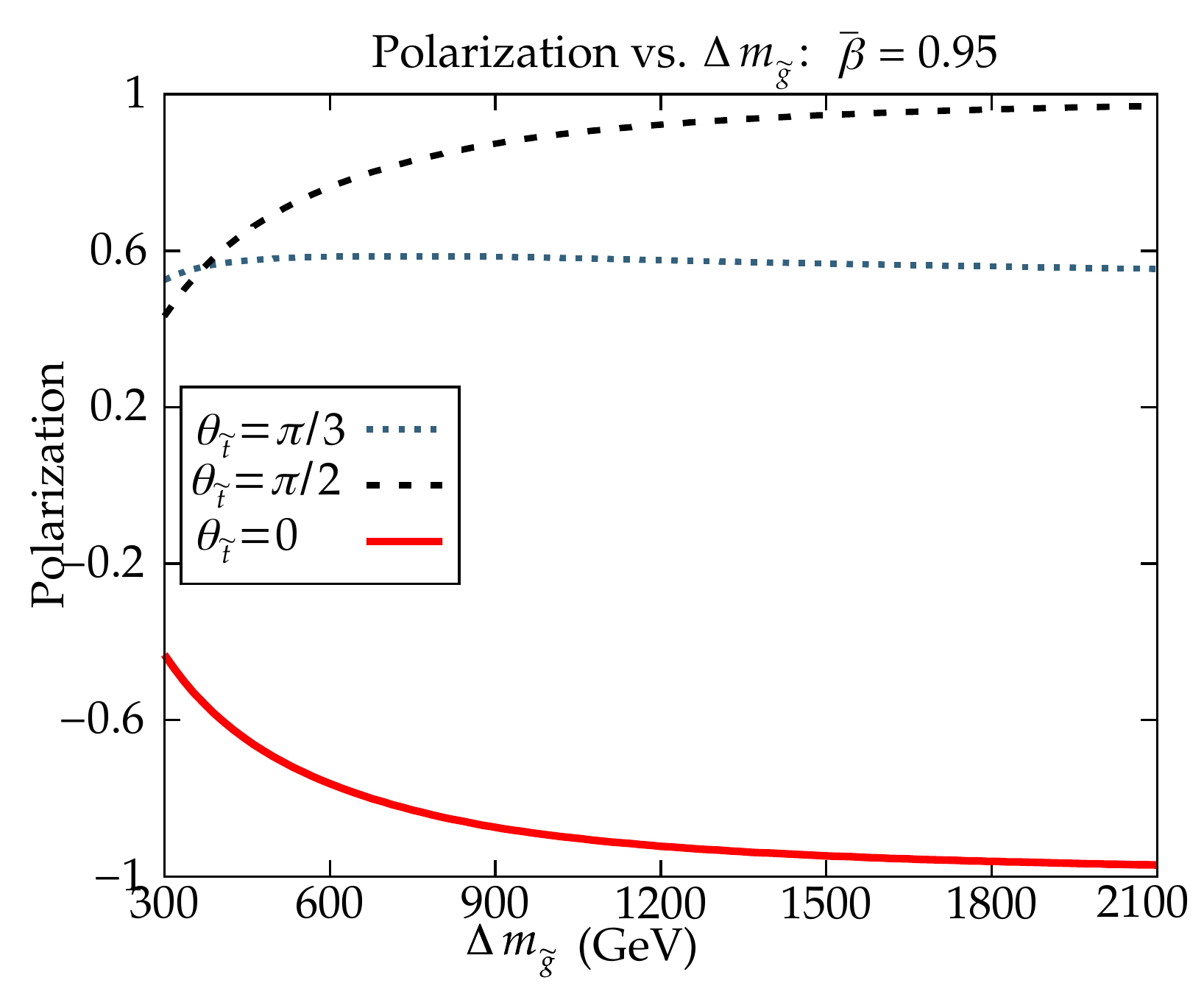}
\caption[Top polarization vs. $\Delta m_{\tilde{g}}=m_{\tilde{g}}-m_{\tilde{t}_1}$]{The top polarization as a function of $\Delta m_{\tilde{g}}=m_{\tilde{g}}-m_{\tilde{t}_1}$, for two different values of the velocity of gluino: $\bar{\beta}=0.0$ (top) and $\bar{\beta}=0.95$ (bottom).}\label{fig:2}
\end{figure}
\begin{figure}
\includegraphics[scale=0.45,keepaspectratio=true]{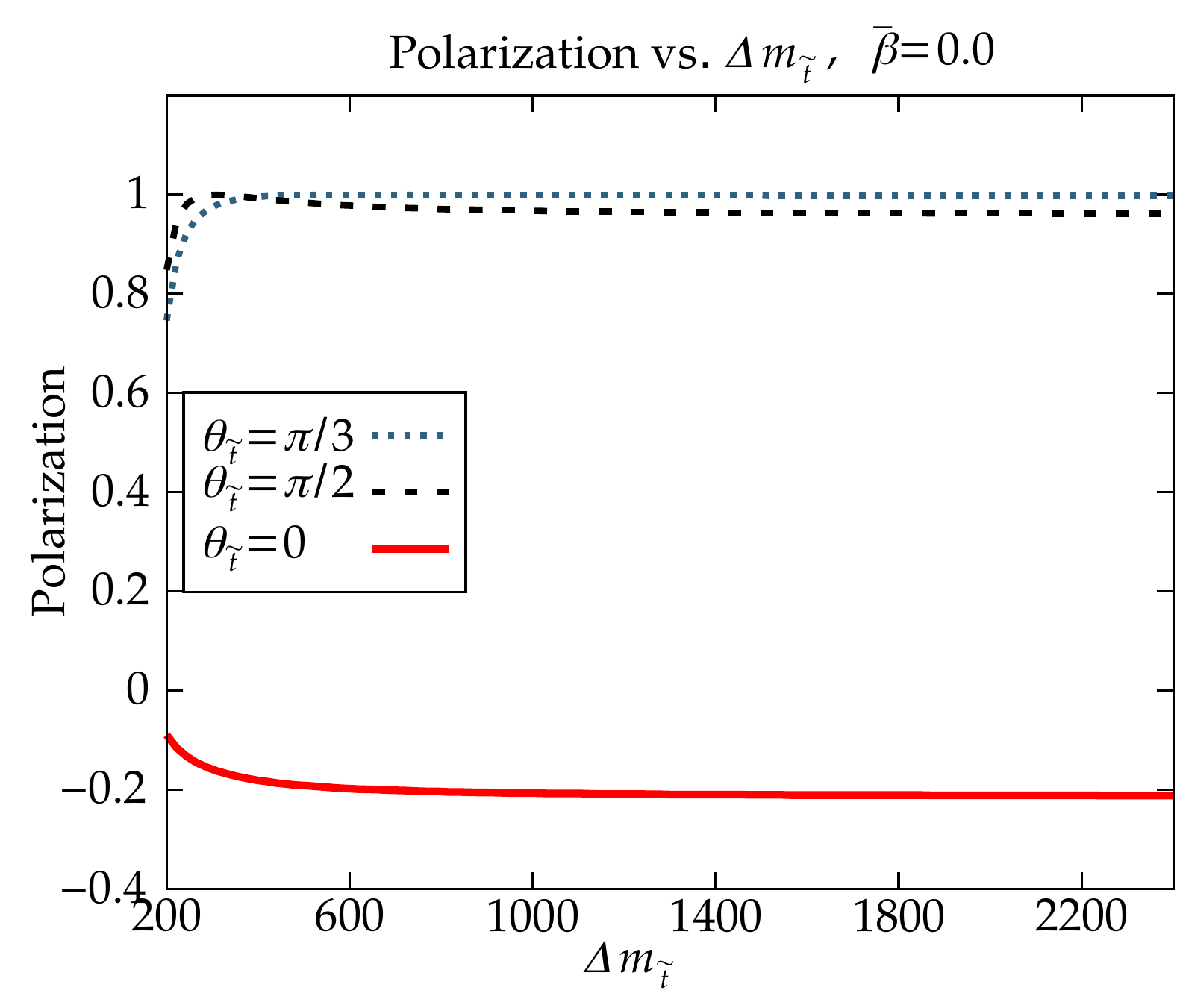}
\includegraphics[scale=0.45,keepaspectratio=true]{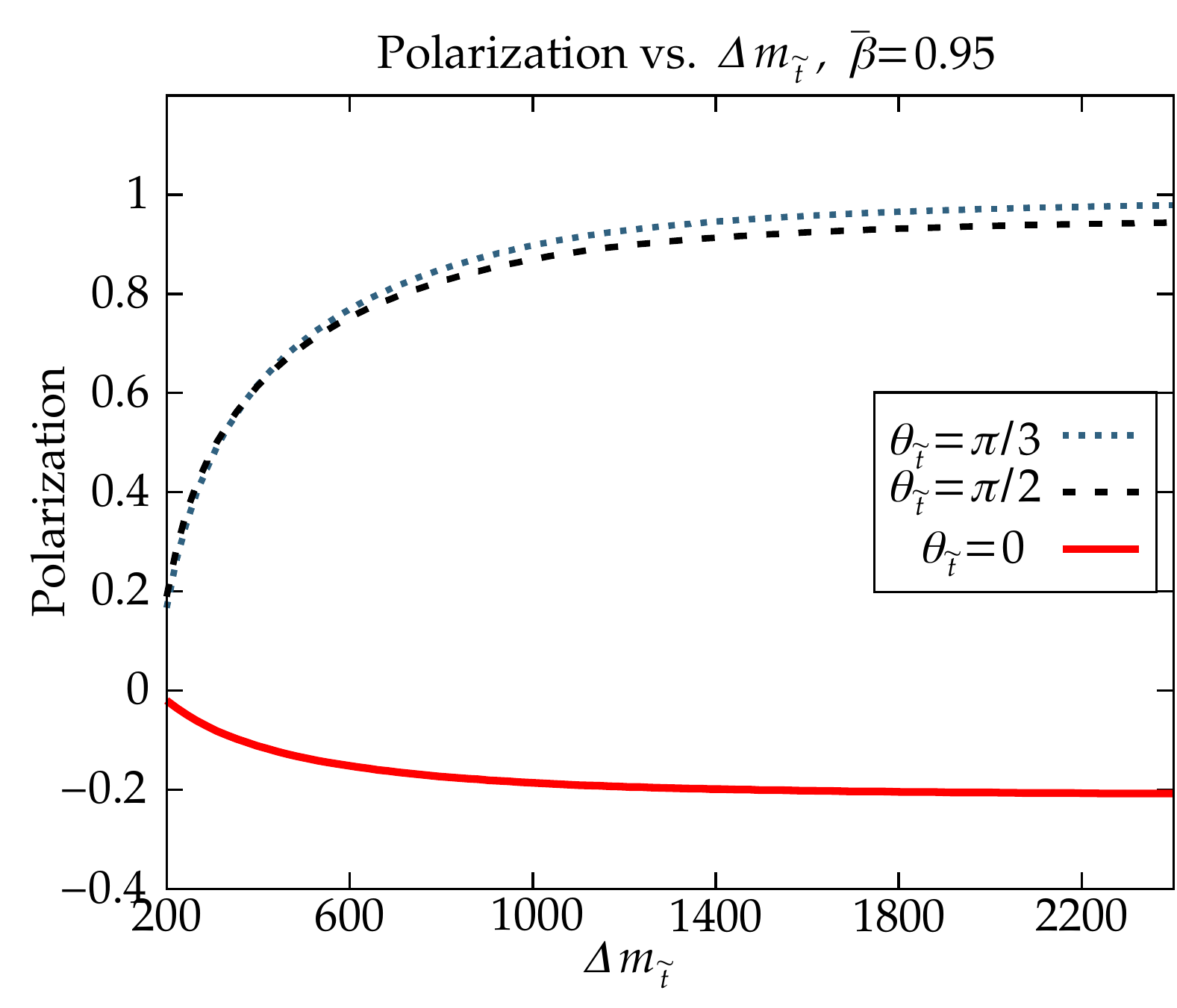}
\caption[Top polarization vs. $\Delta m_{\tilde{t}}=m_{\tilde{t}_1}-m_{\tilde{\chi}^0_1}$]{Top polarization as a function of $\Delta m_{\tilde{t}}=m_{\tilde{t}_1}-m_{\tilde{\chi}^0_1}$ for different values of the velocity of the stop ($\bar{\beta}$): $\bar{\beta}=0.0$ (top) and $\bar{\beta}=0.95$ (bottom). The neutralino are bino-like. $N_{11}\approx 1$, $N_{12},N_{14}\approx 0$, where $N_{ij}(i,j=1,\cdots,4)$ denote the neutralino mixing matrix elements.}\label{fig:4}
\end{figure}
Eqs.~(\ref{eq:5}) and (\ref{eq:7}) show that $\mathcal{P}_0$ is a function of 
$\Delta m_{\tilde{g}/\tilde{t}_1}$ and the mixing angle $\theta_{\tilde{t}}$. 
Thus, $\mathcal{P}$ depends not only on $\bar\beta$ but also on the mass
difference and the mixing angle.
Figs.~\ref{fig:2} and \ref{fig:4} show $\mathcal{P}$ as a function 
of $\Delta m_{\tilde{g}/\tilde{t}}$ for different choices of
$\theta_{\tilde{t}}$ and $\bar\beta$. In the stop case, the neutralino is 
assumed to be bino-like with $N_{11}\approx 1$, $N_{12},N_{14}\approx 0$.
The function $I(\bar{\beta})$ for small values of $\Delta
m_{\tilde{g}/\tilde{t}}$, i.e. for small values of $\beta$ (velocity of the top
in the rest frame of the mother particle) varies as $\beta/\bar{\beta}$
according to Eq.~(\ref{eq:25}). For large values of $\Delta
m_{\tilde{g}/\tilde{t}}$ such that $\beta>\bar{\beta}$, $I(\bar{\beta})$ goes
as $1/\beta^2$ which remains close to unity.  Hence, $I(\bar{\beta})$ in these
cases changes rapidly as a function of $\Delta m_{\tilde{g}/\tilde{t}}$ and
approaches its limiting value asymptotically as $\Delta m_{\tilde{g}/\tilde{t}}$
increases further towards large values.  This explains the rapid rise of the
magnitude of top polarization with increasing $\Delta m$ for
$\theta_{\tilde{t}}=\pi/2$  and $\theta_{\tilde{t}}=0$, in the case of gluino
decay. The case of stop decay is also qualitatively similar, as seen from Fig.~\ref{fig:4}.
\begin{figure}
\includegraphics[scale=0.45,keepaspectratio=true]{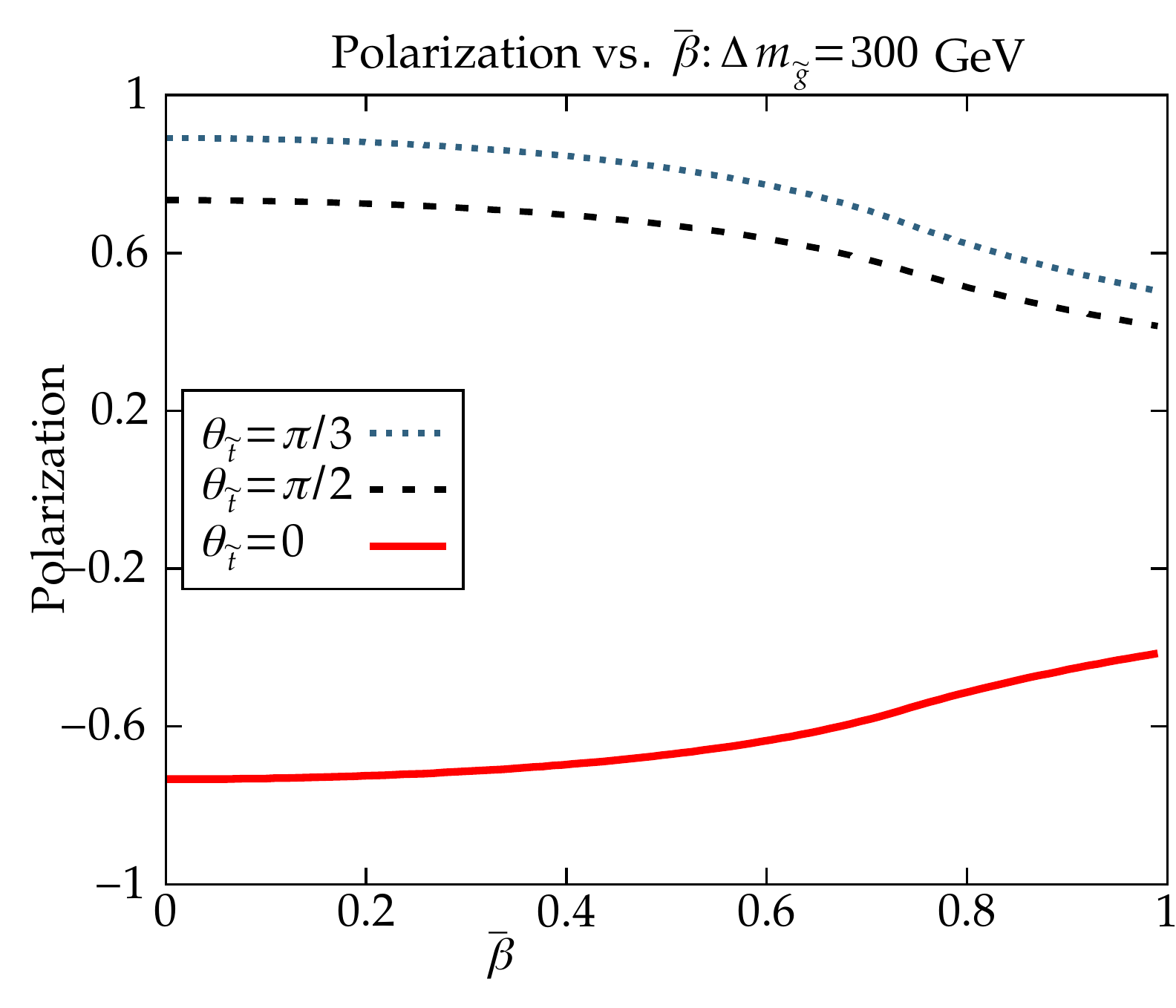}
\includegraphics[scale=0.45,keepaspectratio=true]{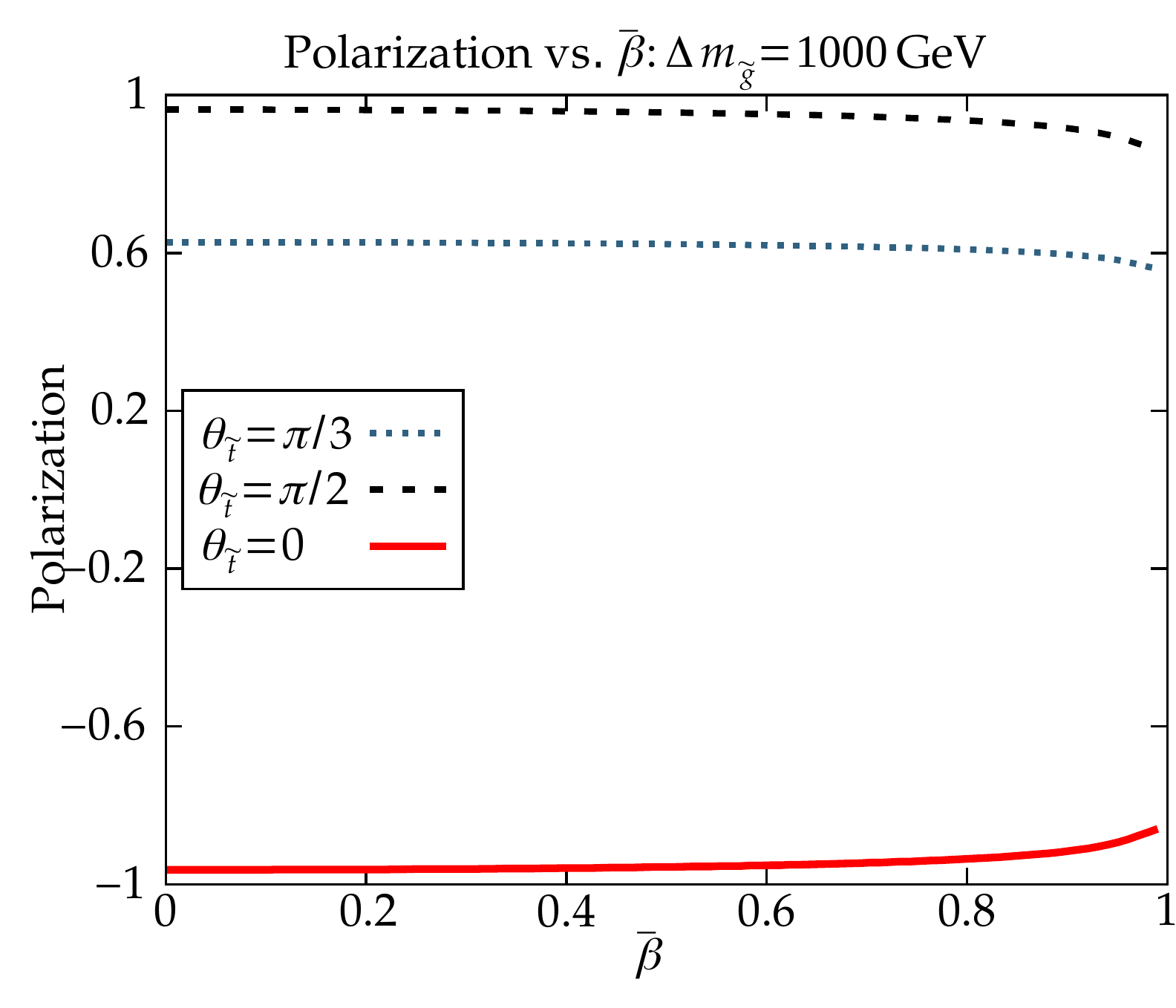}
\caption[Top polarization vs. velocity ($\bar{\beta}$) of gluino]{The top polarization as a function of velocity $\bar{\beta}$ of the gluino, for two different values of $\Delta m_{\tilde{g}}=m_{\tilde{g}}-m_{\tilde{t}_1}$: $\Delta m_{\tilde{g}}=300$ GeV (top) and $\Delta m_{\tilde{g}}=1000$ GeV (bottom).}\label{fig:1}
\end{figure}



\begin{figure}
\includegraphics[scale=0.45,keepaspectratio=true]{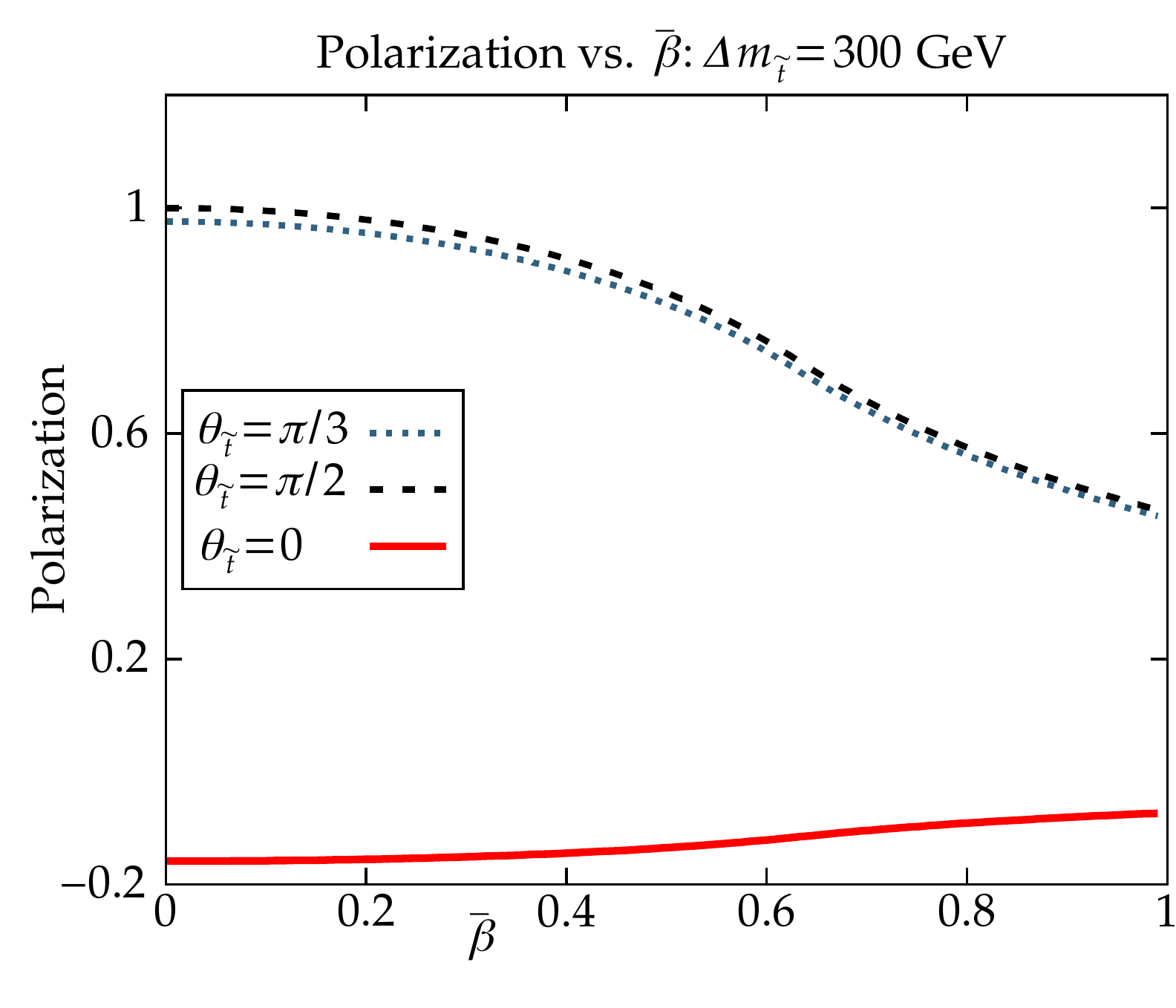}
\includegraphics[scale=0.45,keepaspectratio=true]{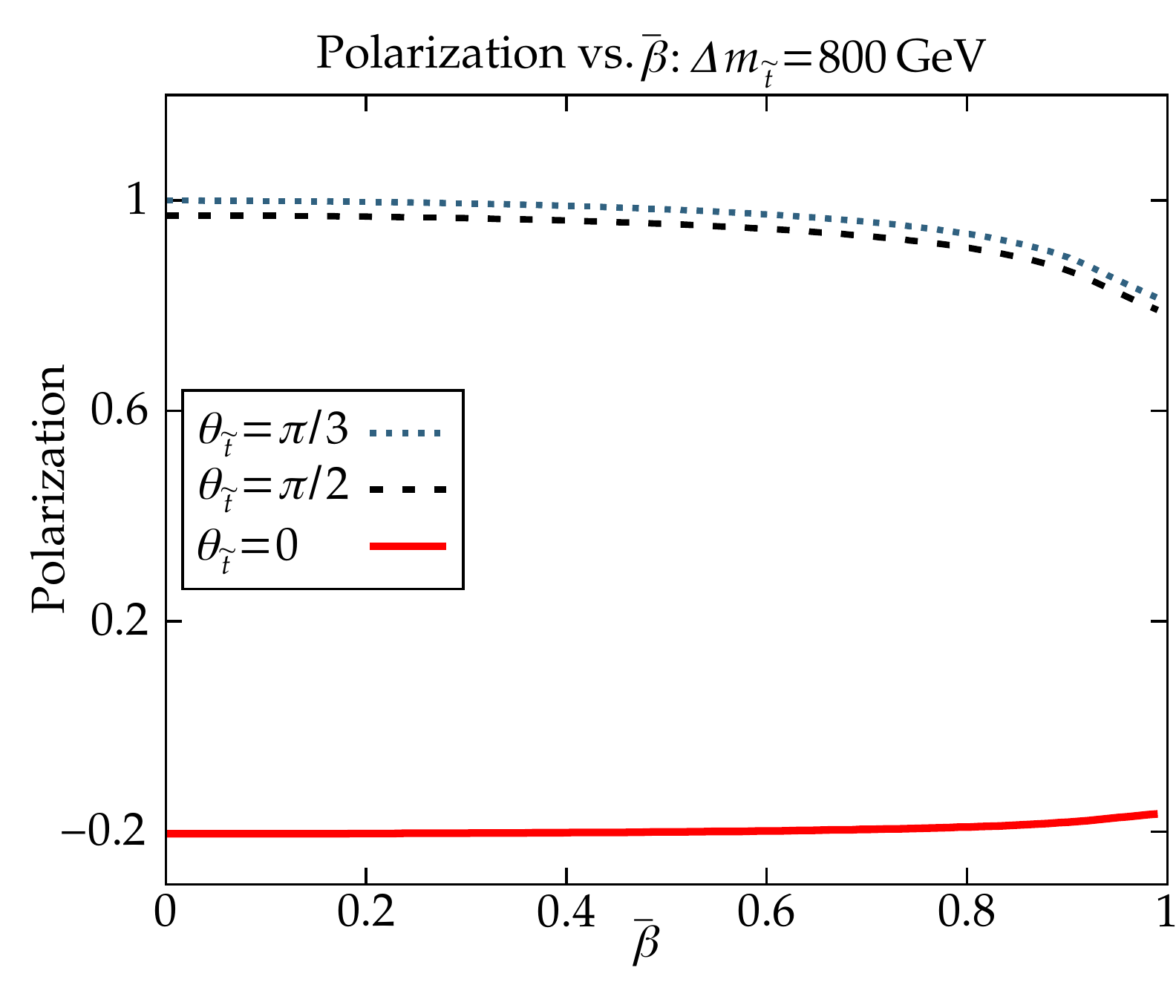}
\caption[Top polarization vs. velocity ($\bar{\beta}$)of the stop.]{Top polarization as a function of the boost $\bar{\beta}$ of the stop, for two different values of the mass difference between the stop and the neutralino ($\Delta m_{\tilde{t}}=m_{\tilde{t}_1}-m_{\tilde{\chi}^0_1}$): $\Delta m_{\tilde{t}}=300$ GeV (top) and $\Delta m_{\tilde{t}}=800$ GeV (bottom). The rest of the parameters are the same as those in Fig.~\ref{fig:4}.}\label{fig:3}
\end{figure}
  
Fig.~\ref{fig:1} (Fig.~\ref{fig:3}) show the top polarization as a function of 
the velocity of the gluino (stop) $\bar{\beta}$ for different choices of 
$\theta_{\tilde{t}}$ and $\Delta m_{\tilde{g}}=m_{\tilde{g}}-m_{\tilde{t}_1}$ 
($\Delta m_{\tilde{t}}=m_{\tilde{t}_1}-m_{\tilde{\chi}_1}$). From 
Eq.~(\ref{eq:25}) we can see that the function $I(\bar{\beta})$ remains 
independent of $\bar{\beta}$ for small values of $\bar{\beta}$ 
($I(\bar{\beta})\propto 1/\beta^2[1-(1-\beta^2)
(1+\mathcal{O}(\bar{\beta}^2))]$) and falls as  $1/\bar{\beta}$ for large 
values of $\bar{\beta}$ ($>\beta$). This can be seen clearly from 
Fig.~\ref{fig:1} and Fig.~\ref{fig:3}. These figures also show that the value 
of $\Delta m_{\tilde{g}/\tilde{t}}$ determines the value of $\bar{\beta}$ at 
which the function $I(\bar{\beta})$ and hence the top polarization 
$\mathcal{P}(\bar{\beta})$ starts to fall as $1/\bar{\beta}$.

Some of the previous works~\cite{Belanger:2012tm,Belanger:2013gha,
Shelton:2008nq} have pointed out the need to consider the effects of kinematics 
on the top polarization when it is measured in the lab 
frame,in cases where the top is produced in the decays of heavy SUSY 
particles. But our work is new in the sense that we have given explicit 
expressions for the top polarization in the case of gluino decay which has not 
been considered in the literature so far. Although Ref.~\cite{Shelton:2008nq} 
has outlined the method of obtaining the top polarization when the top is 
produced in the decays of other particles, we feel that a detailed deviation 
of the expression for top polarization would serve to clarify the issues such 
as the absence of dependence of top polarization on the direction of motion 
of the decaying particle.   

\section{Top polarization in $pp$ collisions}\label{sec:3a}
Note that Eqs.~(\ref{eq:28}) and (\ref{eq:30}) are at parton level. 
The polarization of the top at the level of $pp$ collisions can be obtained by 
convoluting Eq.~(\ref{eq:28}) and Eq.~(\ref{eq:30}) with the parton 
distribution functions of the proton. In the case of gluino decay, this gives the top polarization $\mathcal{P}_{NW}$ in the $pp$ 
collision as a weighted average over the parton center of mass frame 
velocities of the gluino (similar expressions hold for the case of stop decay):
\begin{equation}\label{eq:33}
\mathcal{P}_{NW}=\frac{1}{\sigma_{\tilde{g}\tilde{g}}}\sum_{q_1,q_2}\int\!\!
dx_1 dx_2\; f_{q_1/p}(x_1)f_{q_2/p}(x_2) \ 
\hat{\sigma}_{\tilde{g}\tilde{g}}(\hat{s})\mathcal{P}(\bar{\beta})
\end{equation}
where $\sigma_{\tilde{g}\tilde{g}}$ is the cross section for the production of 
a gluino pair. The total cross section of the full process 
$pp\rightarrow \tilde{g}\tilde{t}_1^{\ast}b\bar{\ell}\nu$ is given by 
\begin{align}\label{eq:31}
\sigma(pp\rightarrow \tilde{g}\tilde{t}_1^{\ast}b\bar{\ell}\nu)&=
\sum_{q_1,q_2}\int\!\!dx_1 dx_2\;f_{q_1/p}(x_1)f_{q_2/p}(x_2)\nonumber\\
&\times \hat{\sigma}_{\tilde{g}\tilde{g}} \ 
\mathcal{B}(\tilde{g}\rightarrow t\tilde{t}_1^{\ast}) \ 
\mathcal{B}(t\rightarrow b\bar{\ell}\nu)\\\nonumber
&=\sigma_{\tilde{g}\tilde{g}} \ 
\mathcal{B}(\tilde{g}\rightarrow t\tilde{t}_1^{\ast}) \ 
\mathcal{B}(t\rightarrow b\bar{\ell}\nu).
\end{align}
Equation~(\ref{eq:33}) can also be written in a different form:
\begin{equation}\label{eq:33a}
\mathcal{P}_{NW}=\frac{1}{\sigma_{\tilde{g}\tilde{g}}}\int \frac{d\sigma}{d\bar{\beta}}\mathcal{P}(\bar{\beta}).
\end{equation}
The top polarization in the PCM and in the lab frame follow the same formula. The top 
polarization in lab frame can be derived without referring to the PCM frame 
by considering 
the transformation mentioned in Section~\ref{sec:3} (see also 
Appendix~\ref{sec:app}) \cite{Shelton:2008nq}. The rotation matrix $R$ of 
Eq.~(\ref{eq:20}) simply becomes, $R=R_y(-\omega)$. The expression for 
$\omega$ the of the same form as the one the corresponding to 
Eq.~(\ref{eq:28}) except that $\bar{\beta}$ is replaced by the velocity of 
the gluino in the lab frame $\beta_{\tilde{g}}^{\mathrm{lab}}$. 
Equation~(\ref{eq:22}) becomes, in this case,
\begin{align}\label{eq:33b}
\frac{\mathrm{d}\hat{\sigma}}{\mathrm{d}x_{\ell}\mathrm{d}\cos\theta_{t\ell}
}&\propto\int\hat{\sigma}_{\tilde{g}\tilde{g}}
\Big(1+\mathcal{P}\cos\omega\cos\theta_{t\ell}\Big)x_{\ell}(1-x_{\ell})\mathrm{d}\Omega_t^{\tilde{g}}.
\end{align}
Following the same steps given in Section~\ref{ssec:2} we get Eq.~(\ref{eq:33a}) with 
$\bar{\beta}\rightarrow \beta_{\tilde{g}}^{\mathrm{lab}}$. 
\begin{figure}
\centering
\includegraphics[scale=0.47,keepaspectratio=true]{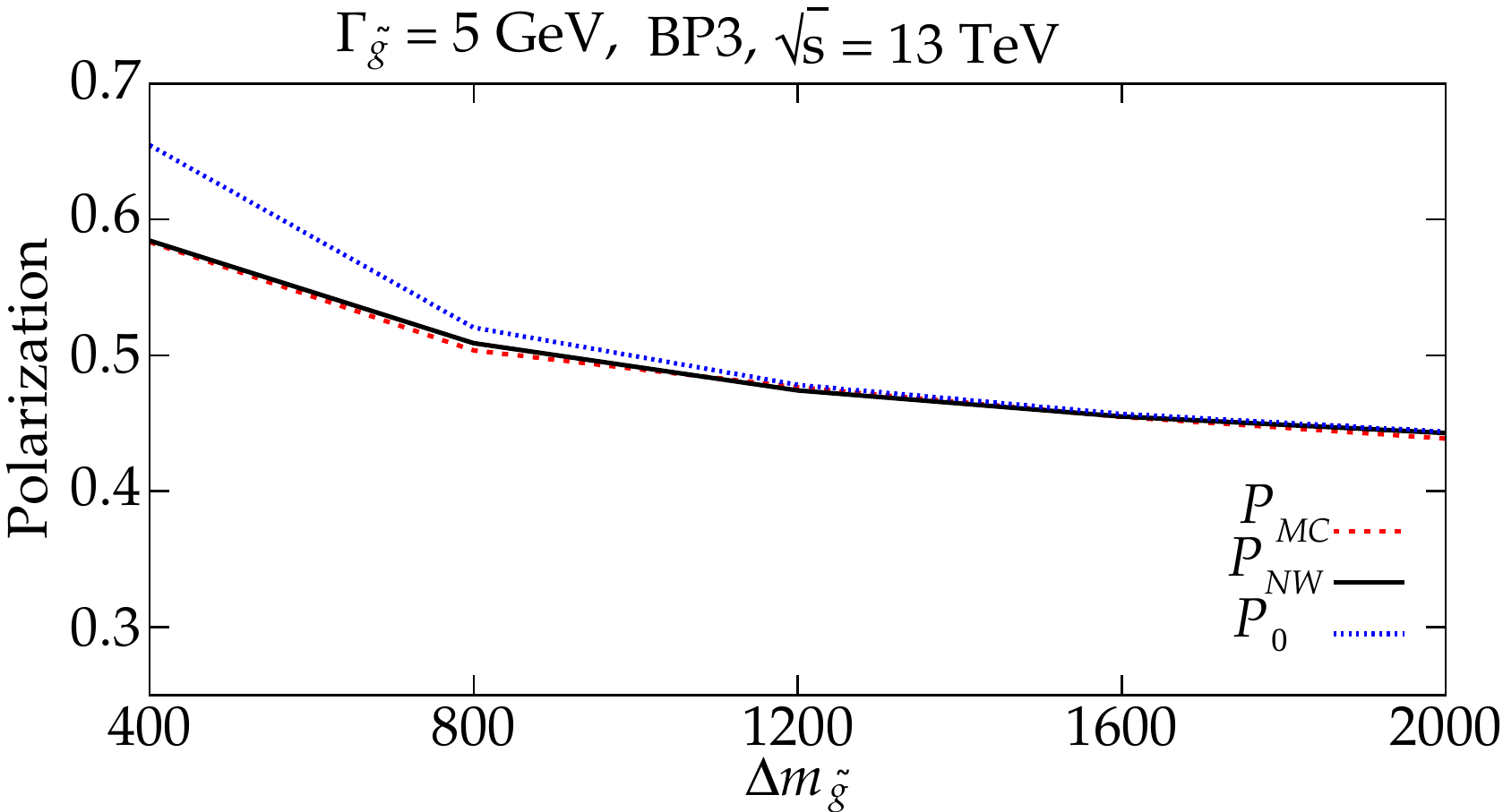}
\includegraphics[scale=0.47,keepaspectratio=true]{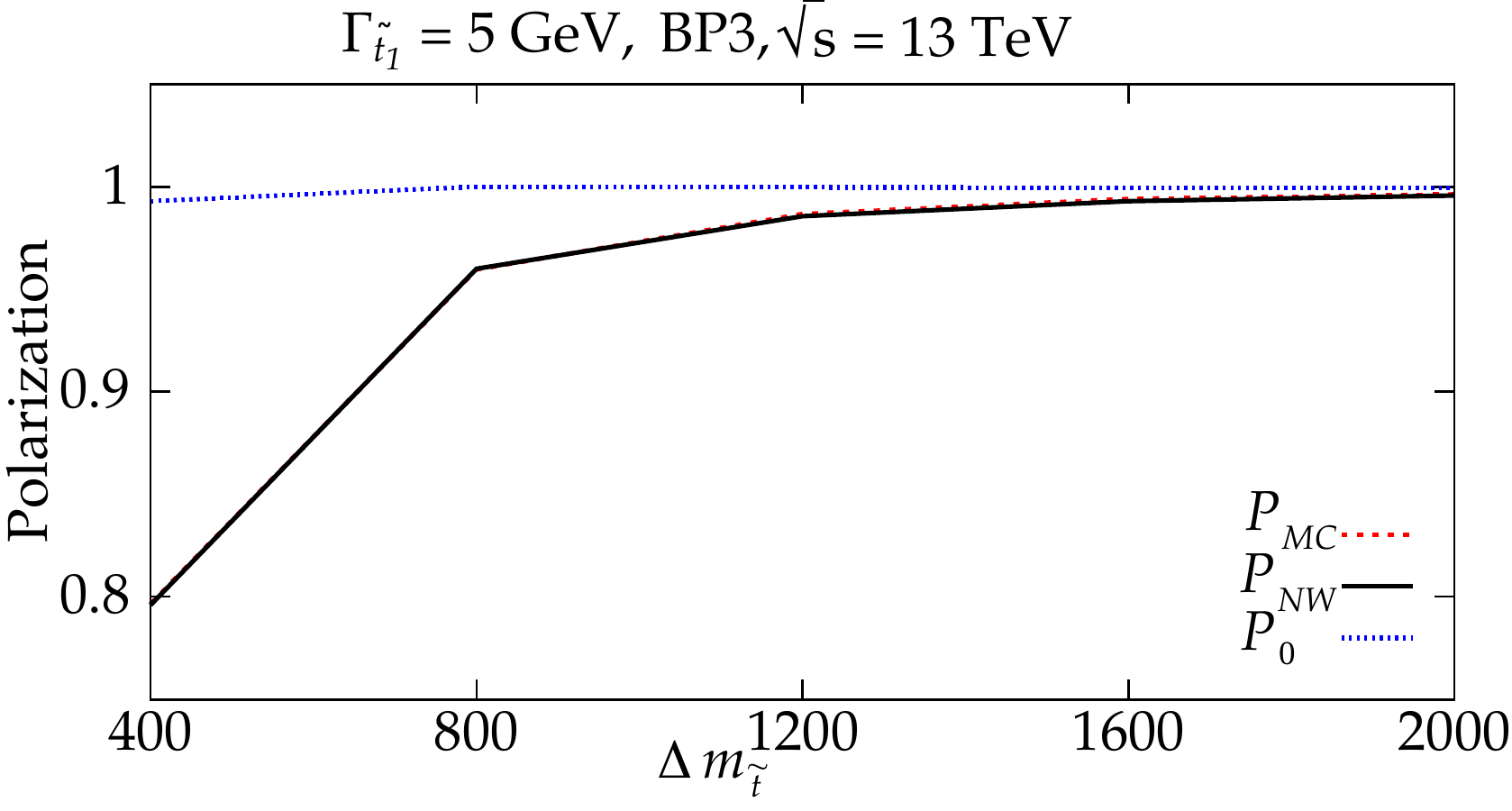}
\caption{Comparison of $\mathcal{P}_{0}$, $\mathcal{P}_{MC}$ and 
$\mathcal{P}_{NW}$ as a function of $\Delta m_{\tilde{g}}$ (top panel) and 
$\Delta m_{\tilde{t}_1}$ (bottom panel) for $\sqrt{s}=13$ TeV LHC, and for the benchmark point BP3. 
The width of the gluino (stop) is taken to be $\Gamma=5$ GeV.}
\label{fig:num1}
\end{figure}

We now present the numerical validation of our method which is summarized by 
Eq.~(\ref{eq:33a}). We use {\tt MadGraph}~\cite{Alwall:2014hca} to generate
events for the processes 
$pp\rightarrow \tilde{g}\tilde{g}\rightarrow t\tilde{t}_1^{\ast}\tilde{g}$ and 
$pp\rightarrow\tilde{t}_1\tilde{t}_1^{\ast}\rightarrow t\tilde{\chi}_1^{0}$, 
followed by the decay of the top through $t\rightarrow b\bar{\ell}\nu$,  for 
the three SUSY benchmark points listed in Table. \ref{tab:1}. 
In these simulations, we artificially vary the 
parameters like width, mass etc. but keep the mixing of stop, and neutralino as 
constants. 

We evaluate the top polarization using different methods for comparisons. 
Firstly, the top polarization is directly obtained from the MC simulation of 
the entire decay chain and then using the formula \cite{Boudjema:2009fz}:
\begin{equation}\label{eq:34}
\mathcal{P}_{MC}=2 \ \frac{N(p_{\ell}\cdot s_t<0)-N(p_{\ell}\cdot
s_t>0)}{N(p_{\ell}\cdot s_t<0)+N(p_{\ell}\cdot s_t>0)}
\end{equation}
where $p_{\ell}$ is the momentum of the charged lepton from the top decay and 
$s_t$ is the longitudinal spin vector satisfying $p_t\cdot s_t=0$ and 
$s_t\cdot s_t=-1$. The top spin vector is defined in the frame in which the 
polarization of the top is defined i.e., the frame of the chosen spin 
quantization axis of the top. 

Secondly, we use the convolution of Eq.~(\ref{eq:28}) or Eq.~(\ref{eq:30}) with 
the distribution of $\bar{\beta}$ obtained from the same MC simulation to
obtain $\mathcal{P}_{NW}$, Eq.~(\ref{eq:33a}). 
Note that the analytical expressions Eq.~(\ref{eq:28}) and (\ref{eq:30}) assume 
the validity of NWA and hence use on-shell mass of the other particle.  This 
estimator gives an average of top polarization, weighted by 
the cross section, over the events of a simulation. 

We first set the width of the decaying gluino/stop to $\Gamma =5$ GeV,
justifying NWA. We expect that the result of $\mathcal{P}_{MC}$ and 
$\mathcal{P}_{NW}$ should agree with each other. This is indeed shown by the 
top and bottom  panels of Fig.~\ref{fig:num1} which compare these two methods 
in the gluino and the stop cases, respectively, as a function of 
$\Delta m_{\tilde{g}}$ or $\Delta m_{\tilde{t}}$. These figures correspond to 
a $pp$ center of mass energy $\sqrt{s}=13$ TeV and the benchmark point BP3.  
The value of rest frame top polarization $\mathcal{P}_0$ is also shown as a 
function of $\Delta m_{\tilde{g}}$ or $\Delta m_{\tilde{t}}$ for comparison. 
We can see that  $\mathcal{P}_{MC}$ and $\mathcal{P}_{NW}$ converges to 
$\mathcal{P}_0$ at large values of $\Delta m_{\tilde{g}/\tilde{t}_1}$. 
This is expected, since the top is highly
boosted in the rest frame of the gluino/stop when $\Delta m$ is large, i.e. 
$\beta\approx 1$. Any boost of this frame $\bar{\beta}$ which is less than 
$\beta$ does not affect the value of top  polarization, since 
$I(\bar{\beta})\approx 1$ with $\bar{\beta}\ll\beta$,  Eq.(\ref{eq:25})). This can be understood physically in the following way: in this limit, due to its large mass, gluino is produced with a small velocity $\bar{\beta}\approx 0$. Hence, we expect that the top polarization has to agree with its value in the gluino rest frame. We do not show the corresponding figures for the other two benchmark points 
as they do show the same good agreement between $\mathcal{P}_{MC}$ 
and $\mathcal{P}_{NW}$ and convergence to $\mathcal{P}_0$ at large values of 
$\Delta m_{\tilde{g}}$ or $\Delta m_{\tilde{t}}$.
\begin{figure}
\centering
\includegraphics[scale=0.47,keepaspectratio=true]{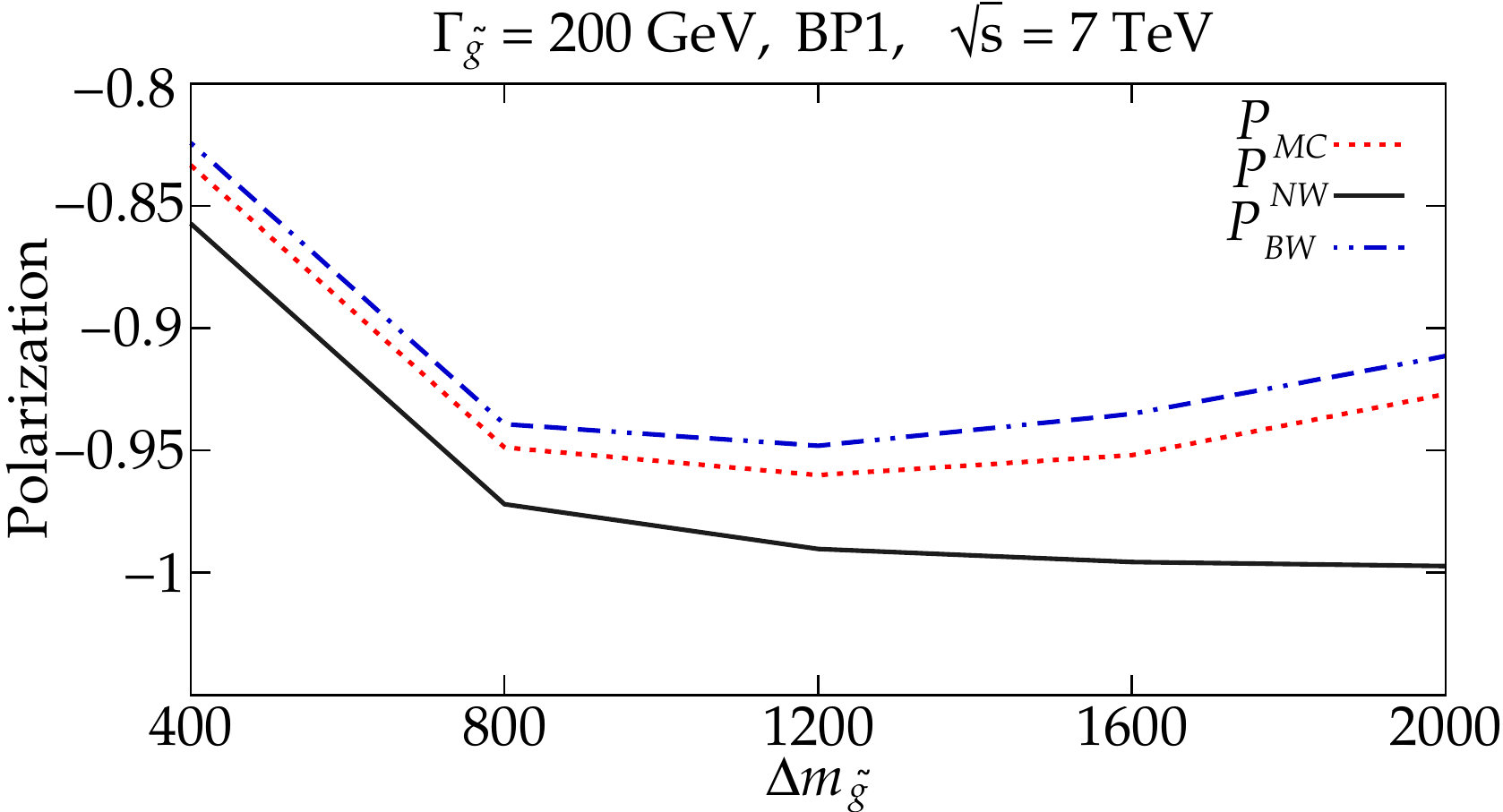}
\includegraphics[scale=0.47,keepaspectratio=true]{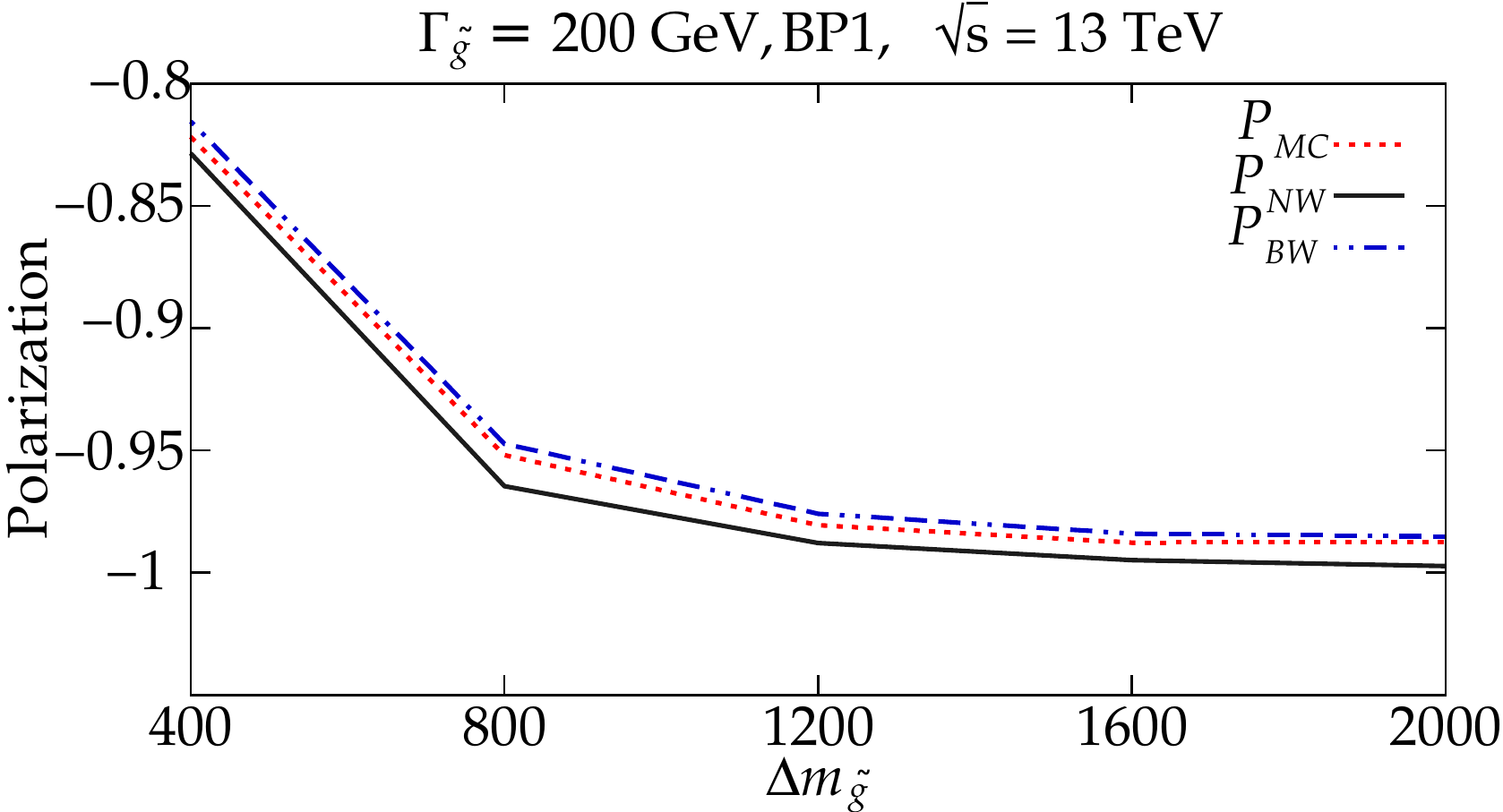}
\caption{Comparison of top polarization, in the case of gluino decay, evaluated in the three methods: $\mathcal{P}_{MC}$, $\mathcal{P}_{NW}$ and $\mathcal{P}_{BW}$, for $\sqrt{s}=7$ TeV(top) and $\sqrt{s}=13$ TeV(bottom), as a function of $\Delta m_{\tilde{g}}$. The width of the gluino is taken to be $\Gamma=200$ GeV. The parameters other than $m_{\tilde{g}},\Gamma_{\tilde{g}}$ correspond to the benchmark point BP1.}
\label{fig:num2}
\end{figure}
\begin{figure}
\centering
\includegraphics[scale=0.47,keepaspectratio=true]{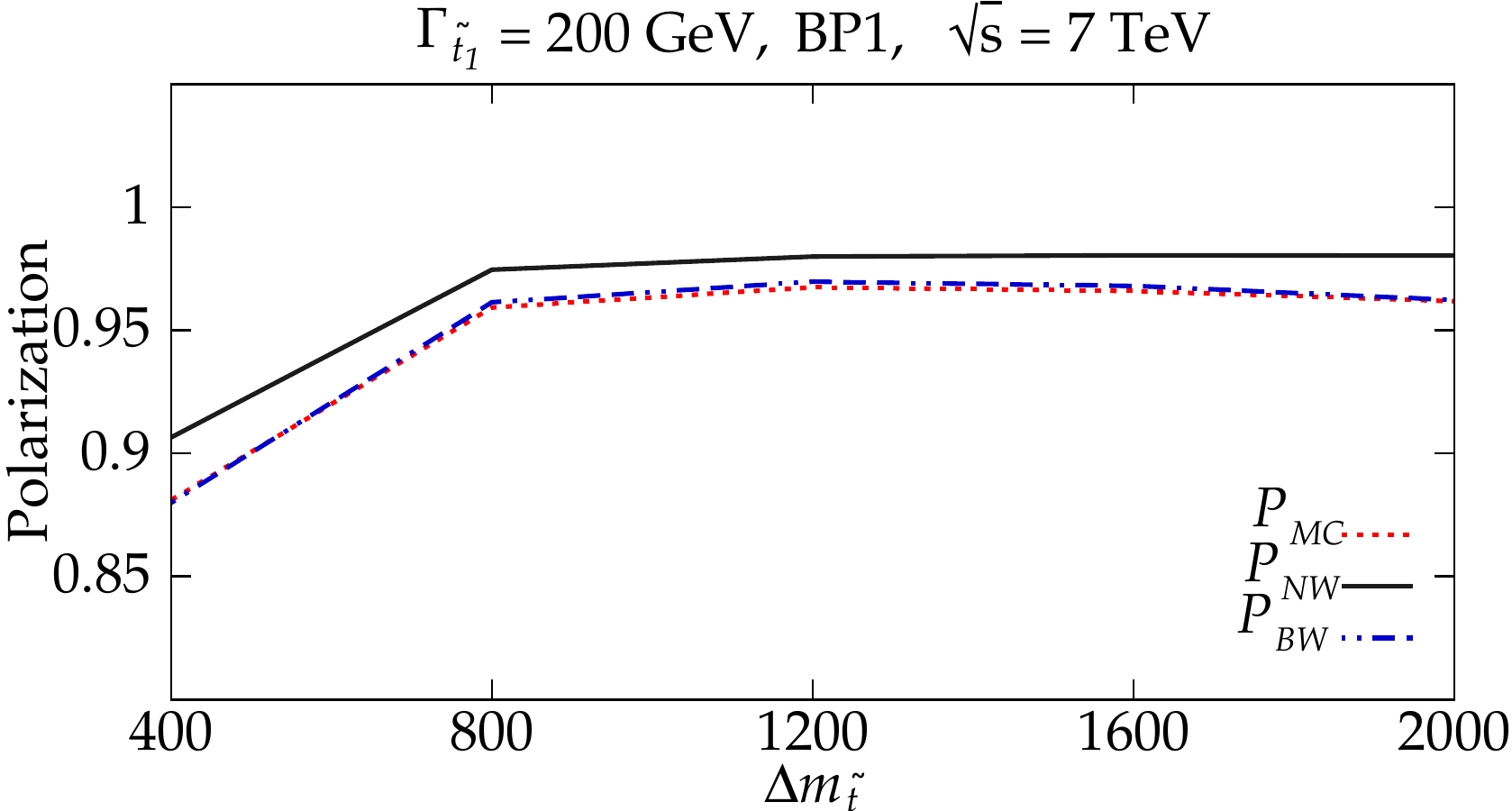}
\includegraphics[scale=0.47,keepaspectratio=true]{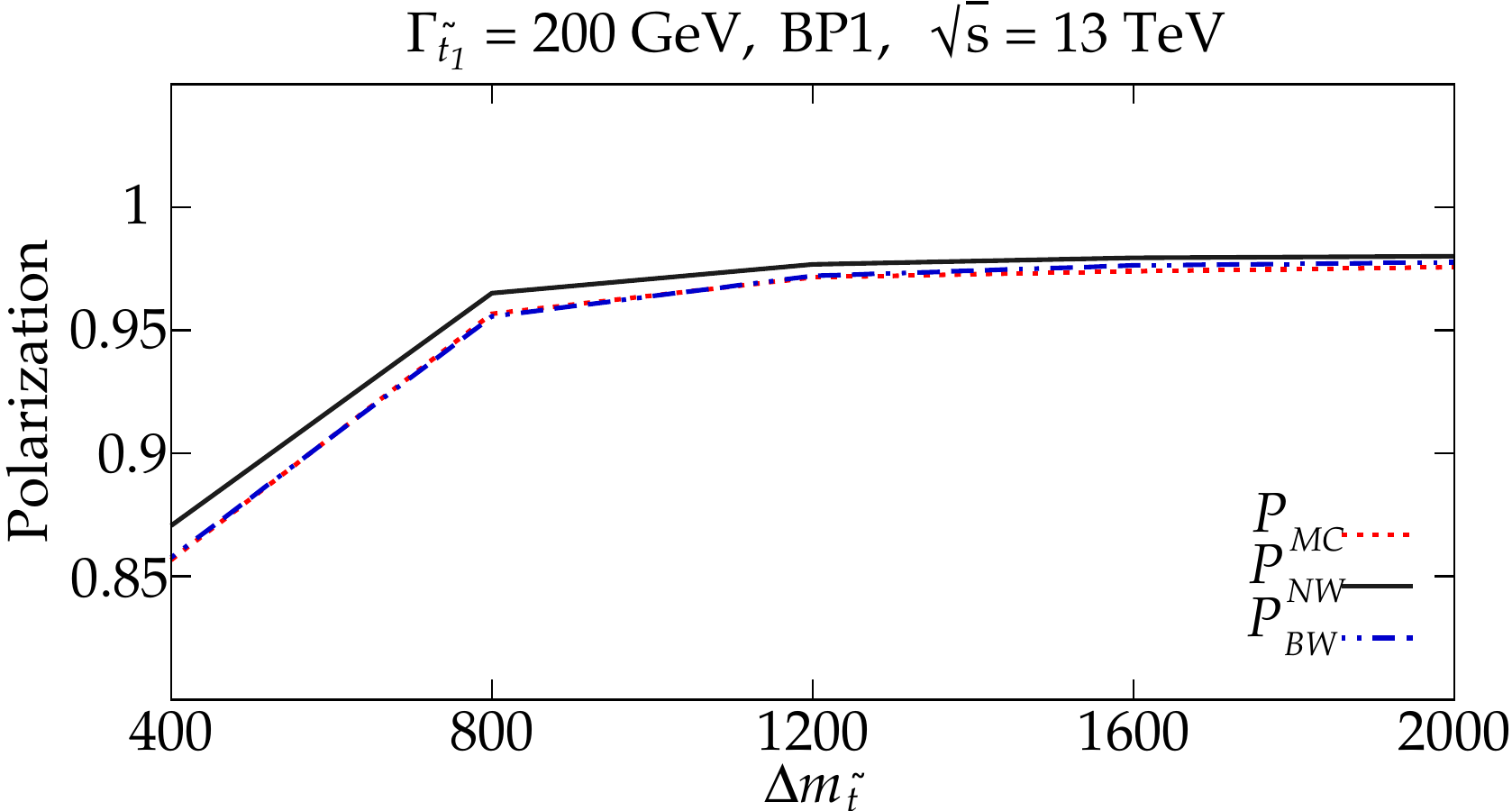}
\caption{Comparison of top polarization, in the case of stop decay, evaluated in the three methods: $\mathcal{P}_{MC}$, $\mathcal{P}_{NW}$ and $\mathcal{P}_{BW}$, for $\sqrt{s}=7$ TeV(top) and $\sqrt{s}=13$ TeV(bottom), as a function of $\Delta m_{\tilde{t}_1}$. The width of the stop and the other parameters are the same as those of Fig.~\ref{fig:num2}.}
\label{fig:num2a}
\end{figure}

As an illustration of the case where the decaying gluino/stop has a finite 
width, we show in Fig.~\ref{fig:num2} (Fig.~\ref{fig:num2a}) the comparison of 
values of top polarization obtained through  $\mathcal{P}_{MC}$ and 
$\mathcal{P}_{NW}$ for the case where the gluino (stop) has a width 
$\Gamma =200$ GeV. In each figure, we present results for two $pp$ center of 
mass energies $\sqrt{s}=7$ TeV and $\sqrt{s}=13$ TeV and for a benchmark point 
BP1. For the range of gluino/stop masses which are considered here, the mother 
particle (gluino/stop) is mostly off-shell when $\sqrt{s}=7$ TeV and mostly 
on-shell for $\sqrt{s}=13$ TeV. Hence, we expect that the results of 
$\mathcal{P}_{NW}$ may show deviation with those of $\mathcal{P}_{MC}$ 
for the case $\sqrt{s}=7$ TeV, and expect better agreement 
between the two methods for $\sqrt{s}=13$ TeV. The figures Fig.~\ref{fig:num2} 
and Fig.~\ref{fig:num2a} show that this is indeed the case. We emphasize here 
that our method $\mathcal{P}_{NW}$ is only an approximation which should work 
when the NWA for the mother particle is applicable. However, we can modify 
$\mathcal{P}_{NW}$ to include, at least partly, the effects which arise from 
finite width of the mother particle through a procedure which is explained in the following section.     

\section{Inclusion of finite width effects}\label{sec:4}
The validity of narrow-width-approximation, with reference to BSM physics, has
been a subject of careful investigation~\cite{Berdine:2007uv,Uhlemann:2008pm,
Kauer:2007zc}. We, on the other hand, take a simple minded approach to address
the presence of large widths for the mother particles and test the validity of
our modified estimator, $\mathcal{P}_{BW}$.

This estimator is obtained when the mass of the mother particle is taken to be $M^2_{\tilde{g}}=p_{\tilde{g}}^2 
(M^2_{\tilde{t}_1}=p_{\tilde{t}_1}^2)$, where $p_{\tilde{g}}$ ($p_{\tilde{t}_1}$) is the momentum of the mother, in place of it's on-shell mass $m_{\tilde{g}}$($m_{\tilde{t}_1}$). In addition to this, the invariant mass is assumed to be distributed as a Breit-Wigner-like 
distribution. In other words, the top polarization is obtained by introducing 
an additional convolution over the mass of the decaying heavy particle:
\begin{eqnarray}\label{eq:36}
\mathcal{P}_{BW}&=&\frac{1}{\sigma_{XX}}\int
_{M^2_{\mathrm{min}}}^{M^2_{\mathrm{max}}}dM^2\Delta_{BW}(M,m)\\\nonumber
&&\times \int f_{q_1/p}(x_1)f_{q_2/p}(x_2)\hat{\sigma}_{XX,M}(\hat{s})\mathcal{P}(\bar{\beta}_M)
\end{eqnarray}     
where $\hat{\sigma}_{XX,M}$ ($XX$=$\tilde{g}\tilde{g}$ or $\tilde{t}_1\tilde{t}_1$) and $\bar{\beta}_M$ are evaluated for a gluino or a stop mass of $M$: $\bar{\beta}_M=\sqrt{1-4M^2/\hat{s}}$. 
\begin{eqnarray}\label{eq:37}
\sigma_{XX}&=&\int _{M^2_{\mathrm{min}}}^{M^2_{\mathrm{max}}}
dM^2\Delta_{BW}(M,m)\\\nonumber
&&\times\int f_{q_1/p}(x_1)f_{q_2/p}(x_2)\hat{\sigma}_{XX,M}(\hat{s})
\end{eqnarray}
The Breit-Wigner factor $\Delta_{BW}(M,m)$ is given by
\begin{equation}\label{eq:38}
\Delta_{BW}(M,m)=\frac{1}{(M^2-m^2)^2+M^2\Gamma^2}.
\end{equation}
 The limits of the integration viz., $M_{\mathrm{min}}$ and $M_{\mathrm{max}}$ can be thought of as the minimum and the maximum mass of the off-shell gluino and are specified usually in the form of an integer ($n$) that represents the `distances' of $M_{\mathrm{max},\mathrm{min}}$ from the on-shell mass in units of the width: $M_{\mathrm{min},\mathrm{max}}=m_{\tilde{g}}\pm n\Gamma_{\tilde{g}}$. The equation given above, Eq.~(\ref{eq:36}) can also be written as
\begin{eqnarray}\label{eq:36a}
\mathcal{P}_{BW}&=&\frac{1}{\sigma_{XX}}
\int_{M^2_{\mathrm{min}}}^{M^2_{\mathrm{max}}}dM^2\Delta_{BW}(M,m)\nonumber\\
&&\times \int \frac{d\sigma_{XX}}{d\beta_M}\mathcal{P}(\beta_M)d\beta_M.
\end{eqnarray}
We note that this procedure is, at best, only an approximate one. In the case
of gluino, there are additional spin correlation between the production and
decay of a gluino pair, when the gluino is off-shell
\cite{Vega:1995cc,Ballestrero:1994jn,Richardson:2001df}. Equation ~(\ref{eq:12}), based on which the expressions
Eq.~(\ref{eq:28}) and Eq.~(\ref{eq:30}) have been derived, should be modified to include the off-shell effects:
\begin{equation}\label{eq:39}
\sum_{\lambda}u(p)\bar{u}(p)=\slashed{p}+m+\frac{\slashed{l}}
{2p\cdot l}(m^2-p^2),
\end{equation}
where $l$ is a light-like four-vector and $m$ is the physical on-shell mass \cite{Richardson:2001df}. In the case of stop decay, we naively expect that the inclusion of Breit-Wigner distribution for the mass of the stop as given in Eq.~(\ref{eq:36a}) should be sufficient, as the stop is a scalar. 
However, as we discuss below, even in the case of stop decay, the top polarization calculated using Eq.~(\ref{eq:36a}) can deviate from the actual value, at large values of $\Delta m_{\tilde{t}}$. 
\begin{figure}
\centering
\includegraphics[scale=0.47,keepaspectratio=true]{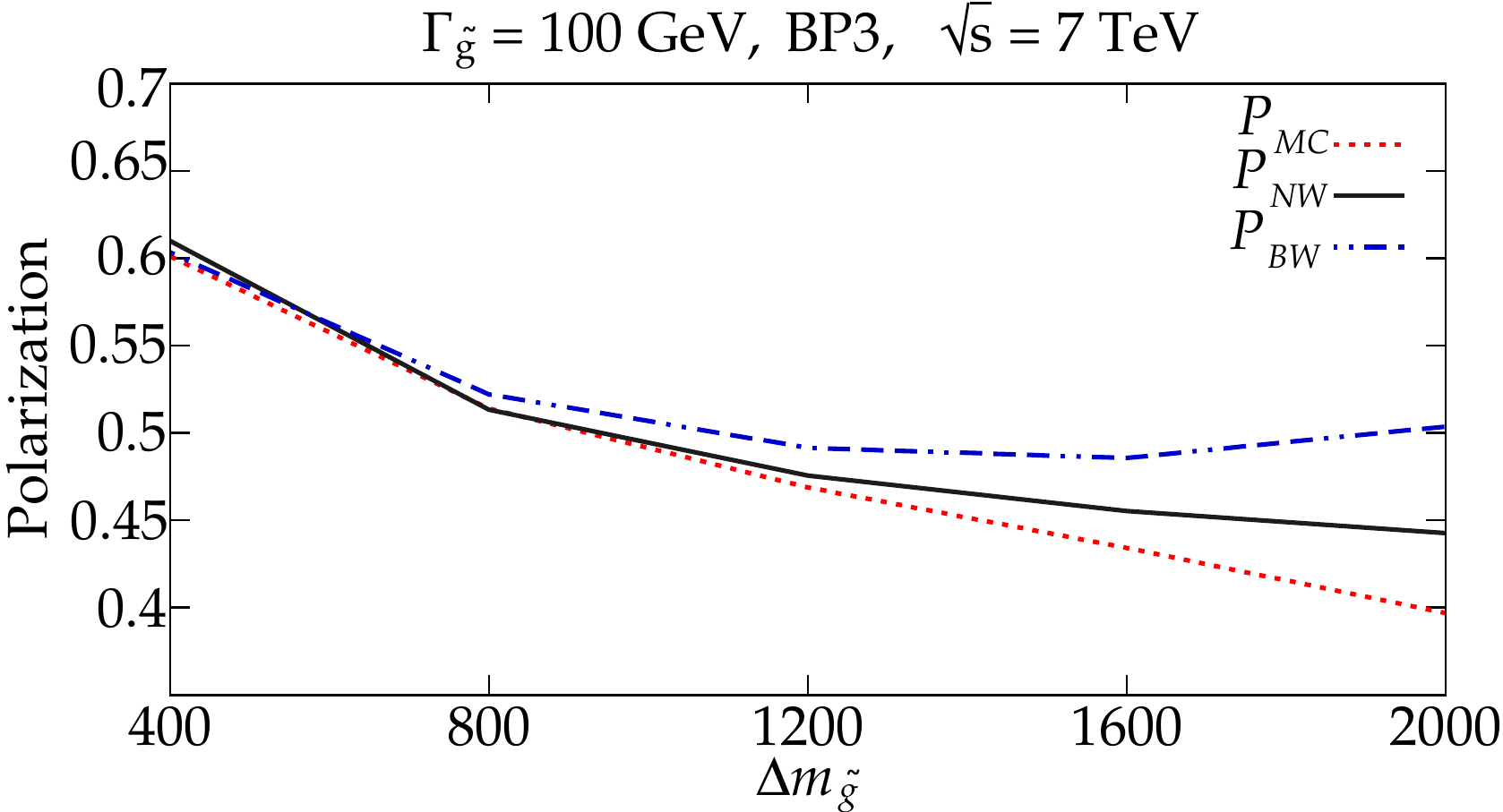}
\includegraphics[scale=0.47,keepaspectratio=true]{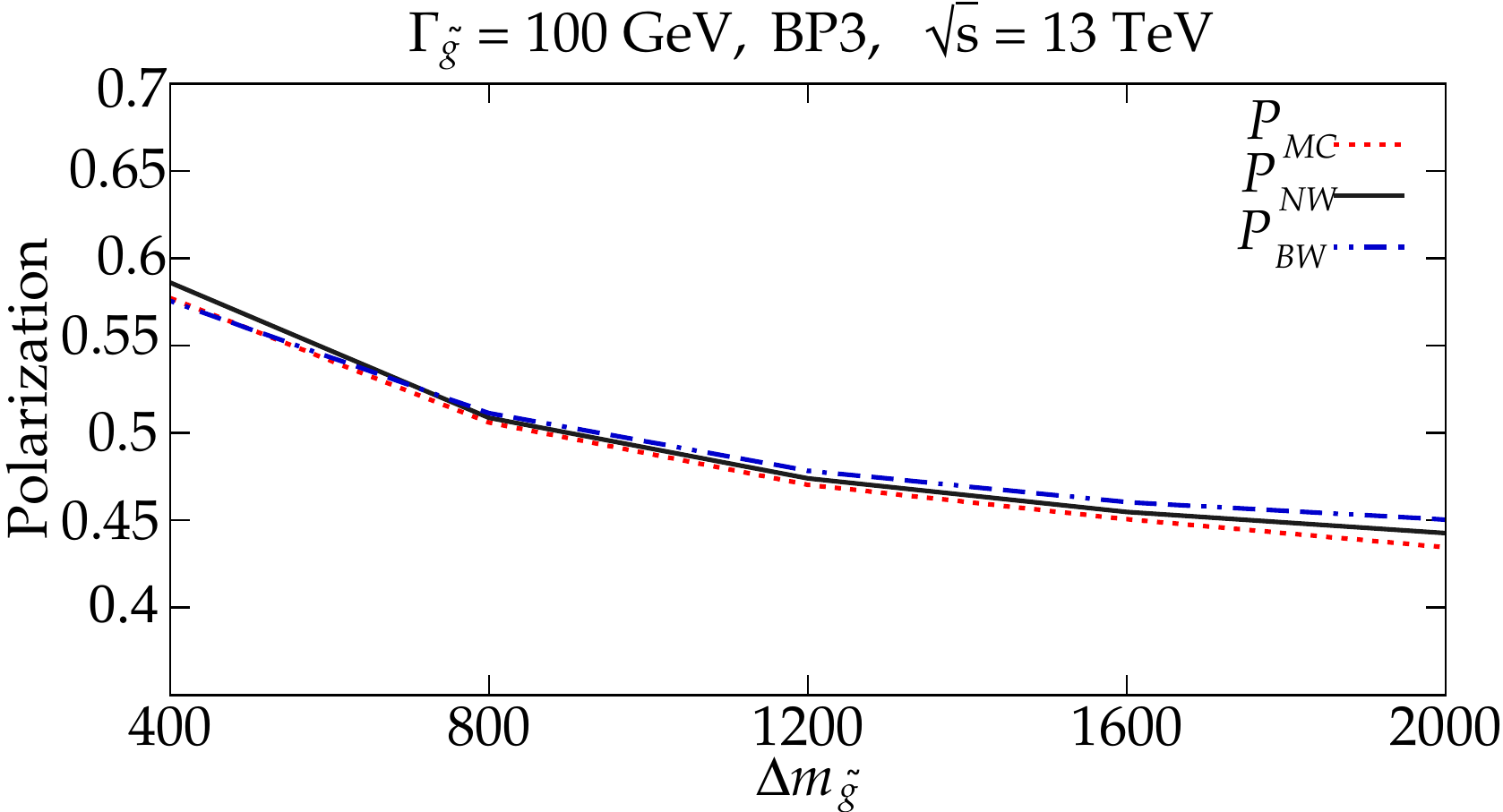}
\caption{Comparison of top polarization evaluated in the three methods: $\mathcal{P}_{MC}$, $\mathcal{P}_{NW}$ and $\mathcal{P}_{BW}$, as a function of $\Delta m_{\tilde{g}}$, for two $pp$ center of mass energies $\sqrt{s}=7$ TeV(top), $\sqrt{s}=13$ TeV(bottom), in the case of gluino decay.  The width of the gluino is taken to be $\Gamma=100$ GeV. The parameters other than $m_{\tilde{g}},\Gamma_{\tilde{g}}$ correspond to the  benchmark point BP3.}
\label{fig:num3}
\end{figure}
\begin{figure}
\centering
\includegraphics[scale=0.47,keepaspectratio=true]{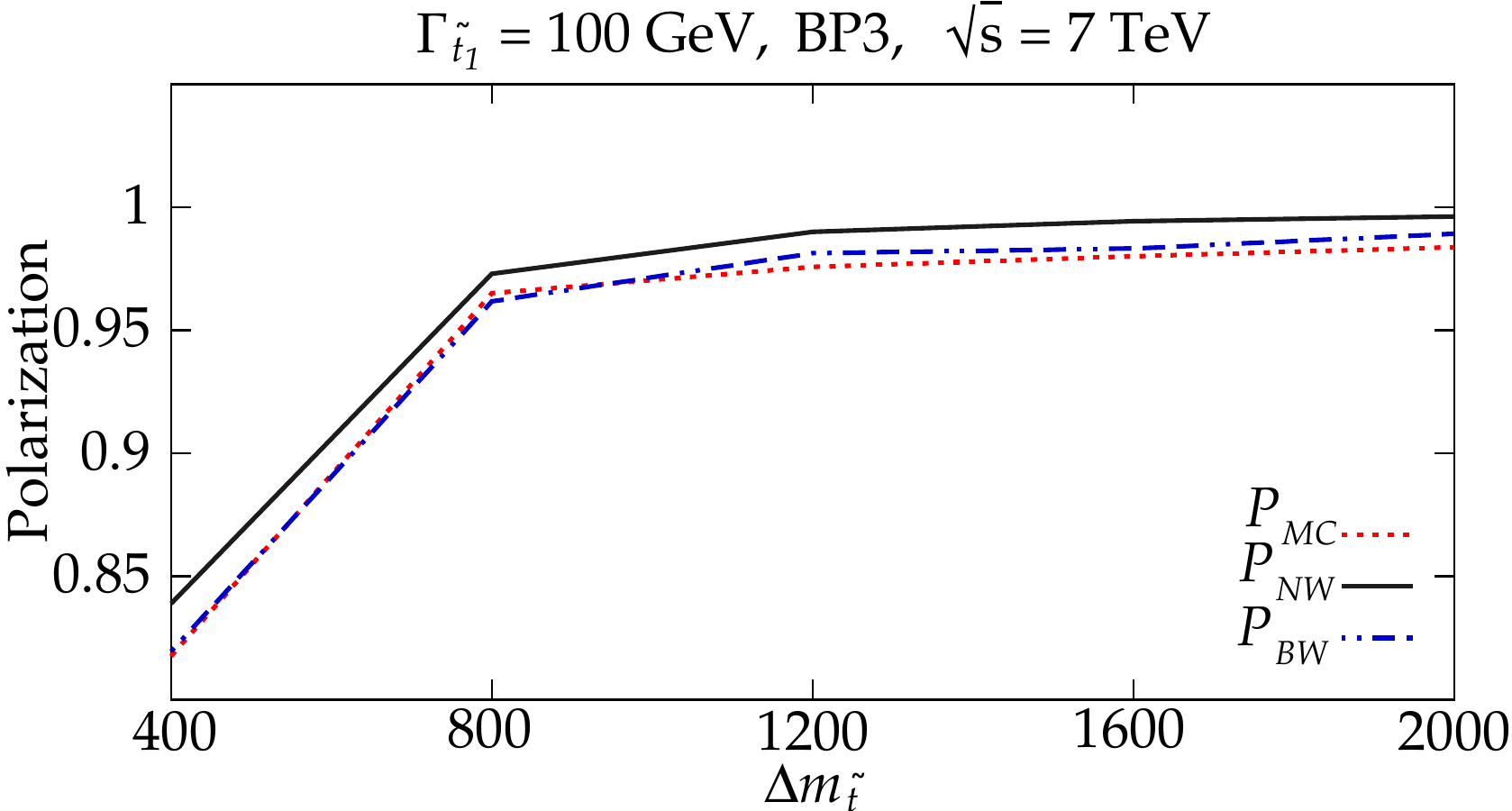}
\includegraphics[scale=0.47,keepaspectratio=true]{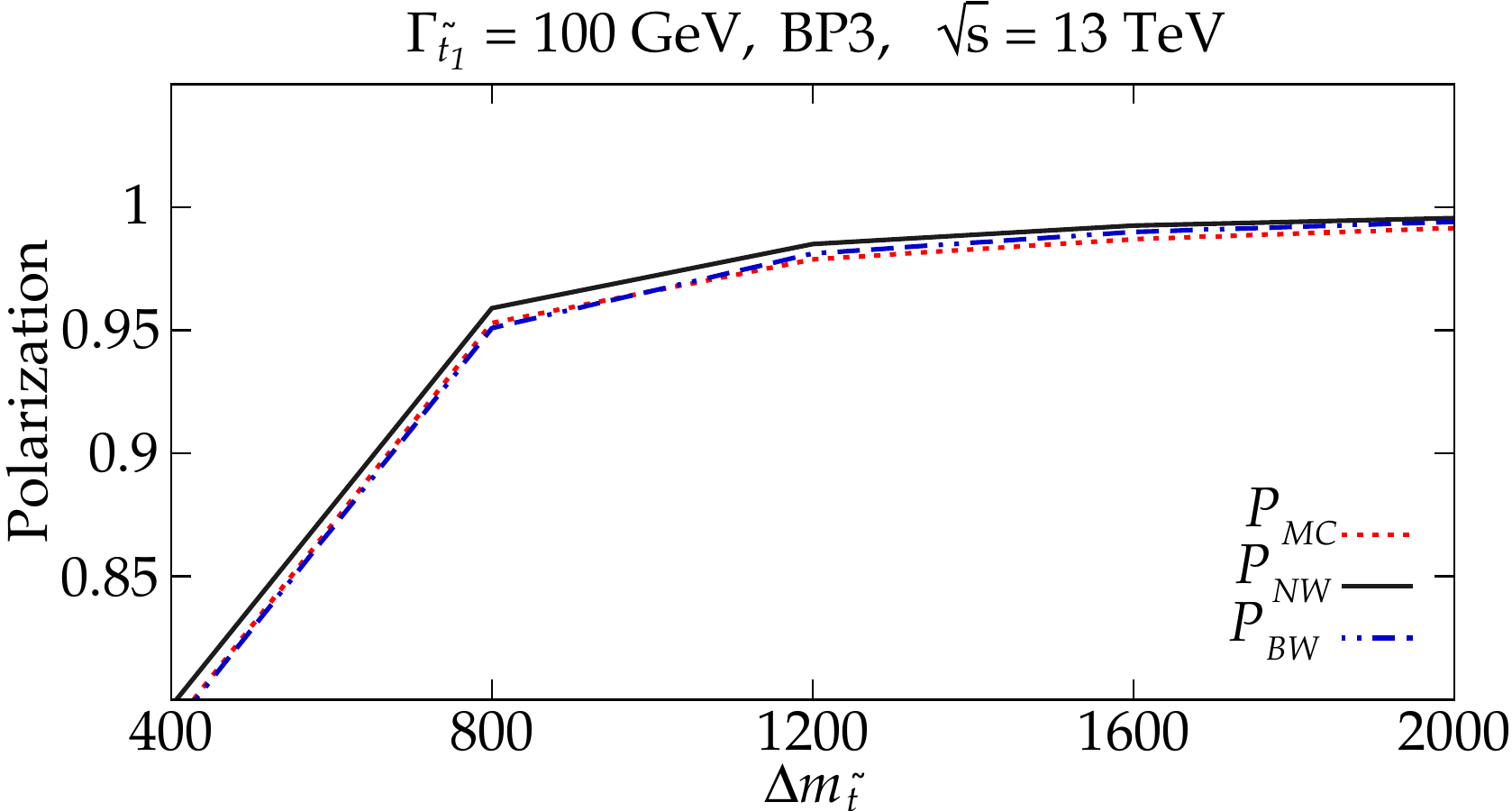}
\caption{Comparison of top polarization evaluated in the three methods: $\mathcal{P}_{MC}$, $\mathcal{P}_{NW}$ and $\mathcal{P}_{BW}$, as a function of $\Delta m_{\tilde{t}}$, for two $pp$ center of mass energies $\sqrt{s}=7$ TeV (top), $\sqrt{s}=13$ TeV(bottom), in the case of stop decay. The width of the stop and the other parameters are the same as those in Fig.~\ref{fig:num3}.}
\label{fig:num4}
\end{figure}
  In the previous section, we have described the polarization $\mathcal{P}_{MC}$
and $\mathcal{P}_{NW}$, as given in Fig.~\ref{fig:num2} and \ref{fig:num2a},
which correspond to cases where the mother particle has a finite width. These
figures also show the polarization $\mathcal{P}_{BW}$, Eq.~(\ref{eq:36a}). 
In the top panels of
Fig.~\ref{fig:num2} and \ref{fig:num2a} which correspond to the case
$\sqrt{s}=7$ TeV, one can clearly see the improvement one obtains in
$\mathcal{P}_{BW}$ over the that of $\mathcal{P}_{NW}$. 
Although $\mathcal{P}_{BW}$ differs from
$\mathcal{P}_{MC}$ particularly for large values of $\Delta m_{\tilde{g}}$ (or
$\Delta m_{\tilde{t}}$) the difference is much smaller compared to the
difference between $\mathcal{P}_{MC}$ and $\mathcal{P}_{NW}$. When the $pp$
center of mass energy $\sqrt{s}$ is increased to 13 TeV, as shown on the bottom
panels of Figs~\ref{fig:num2} and \ref{fig:num2a}, the three methods agree with
each other within a few percent. As mentioned before, this is due to the
increase in the relative contribution of on-shell gluino/stop at $\sqrt{s}=13$
TeV. The deviations between the $\mathcal{P}_{BW}$ and $\mathcal{P}_{MC}$ also
depend upon the benchmark point chosen. In Fig.~\ref{fig:num3} and 
Fig.~\ref{fig:num4} we have shown the comparisons of the three methods for the 
benchmark point BP3. In this case, the inclusion of an additional convolution 
over a Breit-Wigner distribution of mother particle mass, as in 
$\mathcal{P}_{BW}$ does not improve the results of $\mathcal{P}_{NW}$, 
for $\sqrt{s}=7$ TeV. In fact, the difference between $\mathcal{P}_{BW}$ and 
$\mathcal{P}_{MC}$ is greater than that between $\mathcal{P}_{NW}$ and 
$\mathcal{P}_{MC}$.  On the other hand, for $\sqrt{s}=13$ TeV, all the three 
method agree, as they do in the previous cases. Hence, we see that an 
inclusion of a convolution of Breit-Wigner distribution for the mother particle 
mass alone does not always lead to the actual value of top polarization. We 
believe that the neglect of additional spin-correlations in the case of 
gluino, as mentioned before, could be a source of this discrepancy. This 
could  also be the possible reason for the discrepancy between 
$\mathcal{P}_{BW}$ and $\mathcal{P}_{MC}$  being particularly large for the 
case of gluino decay compared to the case of stop decay. In view of the fact 
that for $\sqrt{s}=13$ TeV, all the three results agree, we propose that we 
can stick to $\mathcal{P}_{NW}$ rather than use $\mathcal{P}_{BW}$. In any 
case, our method $\mathcal{P}_{NW}$ is valid only when the contribution of 
on-shell gluino/stop pairs to the cross section dominates over the 
corresponding contribution from the off-shell pairs. In these cases, we have 
already established that $\mathcal{P}_{NW}$ gives a reasonable approximation 
to the actual Monte Carlo value of top polarization. The advantage of 
$\mathcal{P}_{NW}$, though only an approximation, is that it allows for a 
fast estimation of the top polarization, in any frame,  when the velocity 
distribution of the produced mother particle alone is available. Detailed 
simulation of the decay of the mother particle is then not necessary.

  

\section{Summary}\label{sec:5}
In this work, we propose a simple estimator to measure the polarization of 
the top produced in the decays of heavy particles, in any frame, given 
its value in the rest frame of the decaying particle. We quantify the 
kinematical factors that relate the top polarization in the two frames.  
We find that the top polarization in the lab frame depends only on the magnitude
of the velocity and not on the angles of emission  of the mother particle in the
lab frame. The polarization estimators $\mathcal{P}_{NW}$ and 
$\mathcal{P}_{BW}$, in the lab frame, are obtained by convoluting the expression 
for top polarization with the velocity distribution of the mother particle in 
the lab frame. 


The estimator $\mathcal{P}_{NW}$ assumes mother particle to be on-shell and
yields values very close to the true one, $\mathcal{P}_{MC}$, when the mother 
particle has narrow width. For a wider mother particle, we use
$\mathcal{P}_{BW}$ which includes the finite-width effects by a convolution with 
Breit-Wigner distribution of the mass of the mother particle. $\mathcal{P}_{BW}$
works better than $\mathcal{P}_{NW}$ for stop with large width. For a wide 
gluino also $\mathcal{P}_{BW}$ works better than $\mathcal{P}_{NW}$ when 
majority of the events corresponds to the on-shell gluino. 
In the case of a heavy and wide gluino, which is pre-dominantly produced
off-shell, both  $\mathcal{P}_{NW}$ and  $\mathcal{P}_{BW}$ can deviate from 
$\mathcal{P}_{MC}$ by an amount as large as $0.05$ for specific mixing angles.
However, when the mother particle is dominantly produced on-shell, as in the case 
of high $pp$ center of mass energy, these estimators can be used to obtain a fast 
and accurate estimation of the top polarization in the lab frame. 
In the case of gluino decay, we point out that the polarization of the produced 
top can be used as a direct probe of mixing angle in the stop sector, in the 
scenario of a 'natural supersymmetry'. 

\begin{acknowledgments}
One of the authors (APV) would like to thank Gaurav Mendiratta for his constant
encouragements and discussions throughout the duration of the project. He also 
thanks Rafiqul Rahman for his kind help and IISER, Kolkata for hospitality, 
during his stay as a visitor. We also thank Saurabh D. Rindani for taking part 
in the early stages of this project. 
\end{acknowledgments} 
\appendix
\section{Rotation of helicity states under Lorentz boosts}\label{sec:app}
In this appendix, we derive the expressions for the helicity rotation angles
$\omega,\chi$ that are mentioned in the text. We denote the operators
corresponding to rotations and Lorentz transformations by $R$ and $L$. The same
symbols denote the corresponding operations themselves. We denote the helicity
states by $|p,\lambda\rangle$. Under a Lorentz transformation $L$, the helicity
states transform as $|p,\lambda\rangle \rightarrow
L|p,\lambda\rangle=|p',\rangle$, with $p'=L\; p$. The state $|p',\rangle$ is a
state with definite momentum in the Lorentz transformed frame. It does not have
a definte helicity in this frame. Thee reason is that the helicity is not
conserved under a general Lorentz transformation. However, it is conserved as
long as the Lorentz transformation is along the direction of motion of the
particle. It is also invariant under rotations. With this information, we try to
obtain an expression for $|p',\rangle$ in terms of the helicity states
$|p',\lambda'\rangle$ of the new frame. We focus only on the case of a massive
particle.

Consider a helicity state $|p,\lambda\rangle$ of a particle with momentum
$p=(p^0,|\vec{p}|\sin\theta\cos\phi,|\vec{p}|\sin\theta\sin\phi,|\vec{p}|\cos\theta)$
and helicity $\lambda$, in a given frame. The following sequence of
transformation map the helicity  state into a state $|m,s_z=\lambda\rangle$ in
the rest frame of the particle: 
 \begin{equation}\label{eq:app1}
   L_z^{-1}(\beta=|\vec{p}|/p^0)R_y^{-1}(\theta)R_z^{-1}(\phi)|p,\lambda\rangle
=|m,s_z=\lambda\rangle
 \end{equation}
 where $s_z$ denotes the eigenvalue of the $z$-component of spin operator
$\vec{S}$. The sequence of two rotations $R_y^{-1}(\theta)R_z^{-1}(\phi)$ bring
the direction of momentum of the particle to the $z$-axis and the Lorentz
transformation $L_z^{-1}$ takes the resulting state
$||\vec{p}|\hat{z},\lambda\rangle$ to a state in the rest frame of the particle
which we take as the eigenstate of $S_z$ operator. We can also invert this
equation and write helicity state of the particle in terms of
$|m,s_z=\lambda\rangle$:
 \begin{equation}\label{eq:app2}
   |p,\lambda\rangle
R_z(\phi)R_y(\theta)L_z(\beta=|\vec{p}|/p^0)|m,s_z=\lambda\rangle.
 \end{equation}
 We regard this expression as the \textit{definition} of the helicity state of
the particle. For convenience we define the sequence of operations on the right
hand side of the above expression as an operation $h(p)$:
 \begin{equation}\label{eq:app3}
   h(p)\equiv R_z(\phi)R_y(\theta)L_z(\beta=|\vec{p}|/p^0).
 \end{equation}

 We now turn to the case of the top quark produced in the decay of a gluino. The parton center of mass (PCM) frame, as mentioned before, can be reached from the top rest frame ($t$ rest) with the following transformations:
 \begin{equation}\label{eq:app4}
   t\operatorname{-}\mathrm{rest}\xrightarrow{h(p_t^{\operatorname{PCM}})}\mathrm{PCM}
 \end{equation}
where $p_t^{\operatorname{PCM}}$ denotes the momentum of the top in the PCM frame. This transformation maps the states $|m,s_z=\lambda\rangle$ in the top rest frame to the top helicity states in the PCM frame. There is no change in the helicity of the top quark, in this transformation.
\begin{equation}\label{eq:app5}
  |p_t^{\operatorname{PCM}},\lambda\rangle=h(p_t^{\operatorname{PCM}})|m,s_z=\lambda\rangle.
\end{equation}
Now the same PCM frame can also be reached from the top rest frame through the following sequence of transformations:
\begin{equation}\label{eq:app6}
  t\operatorname{-}\mathrm{rest}\xrightarrow{h(p_t^{\tilde{g}})}\tilde{g}\operatorname{-}\mathrm{rest}\xrightarrow{h(p_{\tilde{g}}^{\operatorname{PCM}})}\mathrm{PCM}.
\end{equation}
where $p_t^{\tilde{g}}$ and $p_{\tilde{g}}^{\operatorname{PCM}}$ denote the momenta of the top in the gluino rest frame and the momentum of the gluino in the PCM frame respectively. We denote the velocity and the angles of the gluino in the PCM frame by $\bar{\beta}$, $\theta_{\tilde{g}}$ and  $\phi_{\tilde{g}}$ as in the main text. The above expression means that we first go to the gluino rest frame through the helicity preserving transformation $h(p_t^{\tilde{g}})$ and reach the PCM frame by $h(p_{\tilde{g}}^{\operatorname{PCM}})\equiv R_z(\phi_{\tilde{g}})R_y(\theta_{\tilde{g}})L_z(\bar{\beta})$. Note that in the transformation from the gluino rest frame to the PCM frame, the Lorentz transformation $L_z(\bar{\beta})$ acts along the $z$-direction while the top with $p_t^{\tilde{g}}$ is moving at an angle $\theta$ to the $z$-axis. This Lorentz transformation does not preserve the helicity of the top. The following transformations are just rotations which do not further affect the helicity state of the top. As a result, the helicity state of the top obtained through the transformations of Eq.~(\ref{eq:app5}) and those obtained through Eq.~(\ref{eq:app6}) are different. Now,
\begin{equation}\label{eq:app7}
  h(p_{\tilde{g}}^{\operatorname{PCM}})h(p_t^{\tilde{g}})|m,s_z=\lambda\rangle=|p_t^{\operatorname{PCM}},\rangle
\end{equation}
as in Eq.~(\ref{eq:app1}). Inserting $h(p_t^{\mathrm{PCM}})h^{-1}(p_t^{\operatorname{PCM}})=1$ in Eq.~(\ref{eq:app7}), we get,
\begin{equation}\label{eq:app8}
h(p_t^{\operatorname{PCM}})\left[h^{-1}(p_t^{\operatorname{PCM}})h(p_{\tilde{g}}^{\operatorname{PCM}})h(p_t^{\tilde{g}})\right]|m,s_z=\lambda\rangle=|p_t^{\operatorname{PCM}},\rangle.  
\end{equation}
Now, we can easily see that the terms in $[\cdots]$ correspond to a rotation ($R$) in the rest frame of the top quark, since these set of transformations map $p_t^{t}=(m,\vec{0})$ to itself: $h(p_t^{\tilde{g}})p_t^{t}=p_t^{\tilde{g}}$, $h(p_{\tilde{g}}^{\operatorname{PCM}})p_t^{\tilde{g}}=p_t^{\operatorname{PCM}}$ and $h^{-1}(p_t^{\operatorname{PCM}})p_t^{\operatorname{PCM}}=p_t^{t}=(m,\vec{0})$. Since,
\begin{equation}\label{eq:app9}
  R|m,s_z=\lambda\rangle=R_{\lambda,\lambda'}|m,s_z=\lambda'\rangle
\end{equation}
with $R_{\lambda,\lambda'}$ being the elements of this rotation matrix, we get, from Eq.~(\ref{eq:app8}),
\begin{equation}\label{eq:app10}
  |p_t^{\operatorname{PCM}},\rangle=h(p_t^{\operatorname{PCM}})R_{\lambda,\lambda'}|m,s_z=\lambda'\rangle=|p_t^{\operatorname{PCM}},\lambda'\rangle.
\end{equation}
Hence, the effect of the sequence of transformations $h(p_{\tilde{g}}^{\operatorname{PCM}})$ on the helicity states of top $|p_t^{\tilde{g}},\lambda\rangle$ is equivalent to a rotation in the helicity states of the top in the \textit{transformed} frame. The rotation $R$ is given by the following expression:
\begin{equation}\label{eq:app11}
  R=h^{-1}(p_t^{\mathrm{PCM}})h(p_{\tilde{g}}^{\operatorname{PCM}})h(p_t^{\tilde{g}}) 
\end{equation}
This expression is difficult to evaluate. We can break down this rotation into a product of two rotations by inserting $h(p_t')h^{-1}(p_t')=1$ where $p_t'$ is momentum of the top in an intermediate step in the transformation  from gluino rest frame to the PCM frame: $p_t'=L_z(\bar{\beta})p_t^{\tilde{g}}$. Note that $p_t^{\operatorname{PCM}}=h(p_{\tilde{g}}^{\operatorname{PCM}})p^{\tilde{g}}_t=R_z(\phi_{\tilde{g}})R_y(\theta_{\tilde{g}})L_z(\bar{\beta})p_t^{\tilde{g}}=R_z(\phi_{\tilde{g}})R_y(\theta_{\tilde{g}})p_t'$. With this, the expression for $R$ becomes,
\begin{equation}\label{eq:app12}
  R=\left[h^{-1}(p_t^{\operatorname{PCM}})h(p_t')\right]\left[h^{-1}(p_t')h(p_{\tilde{g}}^{\operatorname{PCM}})h(p_t^{\tilde{g}})\right].
\end{equation}
By a direct computation, we can establish that the two terms correspond to rotations about the $z$-axis and $y$-axis in the top rest frame, respectively, through angles $\chi$ and $\omega$. $R=R_z(\chi)R_y(\omega)$. The expressions for $\chi$ and $\omega$ are given below.  In these expressions, Eq.~(\ref{eq:app13}) and Eq.~(\ref{eq:app14}),  $p_t^{\tilde{g}}$ is given in terms of its velocity $\beta$, and the angles $\theta$ and $\phi$ in the gluino rest frame (the subscripts and the superscripts have been dropped) and the angles of $p_t^{\operatorname{PCM}}$ by $\theta^{\prime\prime}$ and $\phi^{\prime\prime}$. The gluino momentum $p_{\tilde{g}}^{\operatorname{PCM}}$ is given in  terms of its velocity $\bar{\beta}$ and the angles $\bar{\theta}$ and $\bar{\phi}$ in the PCM frame (the subscript and the superscript have been dropped).
\begin{align}\label{eq:app13}
\cos{\omega}&=\frac{(\beta+\bar{\beta}\cos\theta)}{\sqrt{\beta^2+\bar{\beta}^2-\beta^2\bar{\beta}^2\sin^2\theta+2\beta\bar{\beta}\cos\theta}},\\\nonumber
\sin{\omega}&=\frac{\bar{\beta}\sin\theta}{\gamma\sqrt{\beta^2+\bar{\beta}^2-\beta^2\bar{\beta}^2\sin^2\theta+2\beta\bar{\beta}\cos\theta}},
\end{align}
and 
\begin{align}\label{eq:app14}
\cos\chi&=\cos\phi\cos\Delta\phi-\sin\Delta\phi\cos\bar{\theta}\sin\phi,\\\nonumber
\sin\chi&=-\frac{\sin\Delta\phi\sin\bar{\theta}\sqrt{\bar{\beta}^{2}+\beta^2-\bar{\beta}^{2}\beta^2\sin^2\theta+2\bar{\beta}\beta\cos\theta}}{\sqrt{1-\bar{\beta}^2}\beta\sin\theta},
\end{align}
where $\Delta\phi=\bar{\phi}-\phi^{\prime\prime}$. 

The lab frame can be reached from the top rest frame by  following two transformations:
\begin{equation}\label{eq:app16}
t\operatorname{-}\mathrm{rest}\xrightarrow{h(p_t^{\tilde{g}})}\tilde{g}\operatorname{-}\mathrm{rest}\xrightarrow{h(p_{\tilde{g}}^l)\equiv R_z(\phi_{\tilde{g}}^{l})R_y(\theta_{\tilde{g}}^{l})L_z(\beta_{\tilde{g}}^{l})}\mathrm{lab}
\end{equation}    
and the helicity preserving one,
\begin{equation}\label{eq:app17}
t\operatorname{-}\mathrm{rest}\xrightarrow{h(p_t^l)\equiv R_z(\phi_t^{l})R_y(\theta^{l}_t)L_z(\beta_t^{l})}\textrm{lab}.
\end{equation}
In these expressions, the quantities which are defined in the lab frame are denoted by a superscript $l$. The helicity states of the top obtained from $|m,s_z=\lambda\rangle$ in the rest frame of the top through the transformation of Eq.~(\ref{eq:app16}) is different from those obtained through the transformation of Eq.~(\ref{eq:app17}). Using $h^{-1}(p_t^{l})h(p_t^{l})=1$ and following the same steps as in the previous case, we obtain
  \begin{equation}\label{eq:app18}
  h(p_{\tilde{g}}^l) h(p_t^{\tilde{g}})|m,s_z=\lambda\rangle=|p_t^{l},\rangle= R_{\lambda,\lambda'}|p_t^{l},\lambda'\rangle
  \end{equation}
  with $R=h^{-1}(p_t^{l})h(p_{\tilde{g}}^l)h(p_{t}^{\tilde{g}})$. This is of the same form as $R_y=h^{-1}(p_t')h(p_{\tilde{g}}^{\mathrm{PCM}})h(p_t^{\tilde{g}})$ of Eq.~(\ref{eq:app12}). Hence, the expressions for the rotation angle $\omega$ , in this case, can be obtained from Eq.~(\ref{eq:app13}) through the replacement: $\bar{\beta}\rightarrow\bar{\beta}^l$. 

\bibliographystyle{apsrev}
\bibliography{paper3}

\end{document}